\documentclass[lettersize,journal]{IEEEtran}
\usepackage{amsmath,amsfonts}
\usepackage{algorithmic}
\usepackage{algorithm}
\usepackage{array}
\usepackage[caption=false,font=normalsize,labelfont=sf,textfont=sf]{subfig}
\usepackage{textcomp}
\usepackage{stfloats}
\usepackage{url}
\usepackage{verbatim}
\usepackage{graphicx}
\usepackage{cite}
\usepackage{caption}
\usepackage{tabularx}
\usepackage{makecell}
\usepackage{comment}
\usepackage{xcolor}

\begin{document}
\title{5G-Advanced Towards 6G: Past, Present, and Future}
\author{
\IEEEauthorblockA{Wanshi Chen, Xingqin Lin, Juho Lee, Antti Toskala, Shu Sun, Carla Fabiana Chiasserini, and Lingjia Liu}
\thanks{Wanshi Chen is with Qualcomm; Xingqin Lin is with Nvidia; Juho Lee is with Samsung; Antti Toskala is with Nokia;  Shu Sun is with Shanghai Jiao Tong University;  Carla Fabiana Chiasserini is with Politecnico di Torino; and  Lingjia Liu is with Virginia Tech.}
}
\maketitle

\begin{abstract}
Since the start of 5G work in 3GPP in early 2016, tremendous progress has been made in both standardization and commercial deployments. 3GPP is now entering the second phase of 5G standardization, known as \textit{5G-Advanced}, built on the 5G baseline in 3GPP Releases 15, 16, and 17. 3GPP Release 18, the start of 5G-Advanced, includes a diverse set of features that cover both device and network evolutions, providing balanced mobile broadband evolution and further vertical domain expansion and accommodating both immediate and long-term commercial needs. 5G-Advanced will significantly expand 5G capabilities, address many new use cases, transform connectivity experiences, and serve as an essential step in developing mobile communications towards 6G. 
This paper provides a comprehensive overview of the 3GPP 5G-Advanced development, introducing the prominent state-of-the-art technologies investigated in 3GPP and identifying key evolution directions for future research and standardization.
\end{abstract}

\begin{IEEEkeywords}
5G, 5G-Advanced, 6G, MIMO, AI, ML, Full-Duplex, Green Networks, UAV, RIS, XR, Positioning, NTN, IAB
\end{IEEEkeywords}

\section{INTRODUCTION}
\label{sec:intro}

The ever-increasing demands for mobile broadband and expansion to the so-called vertical industries (e.g., automotive, satellite, internet of things) continue driving evolution in standard bodies such as the 3rd generation partnership project (3GPP), most recently evidenced by the successful standardization of the fourth generation (4G) long-term evolution (LTE) and the fifth generation (5G) new radio (NR). In particular, 5G NR has witnessed three releases of standardization, providing a robust framework supporting a comprehensive set of use cases over a wide range of spectrum which can be deployed in various scenarios (indoor or outdoor, macro or small cells, etc.).

The next major step in the evolution of 5G towards the sixth generation (6G) mobile networks is called \textit{5G-Advanced}, a new term approved by 3GPP in April 2021 as a response to a new era of 5G. 
Built on the strong 5G foundation, 5G-Advanced will introduce numerous new capabilities to boost the performance and to enable or expand new use cases and verticals to use 5G technology. At the same time, several forward-looking topics as part of 5G-Advanced will also build the bridge towards 6G technology development.  
\subsection{The Path to 5G}
\label{subsec:5gpath}

When the standardization for LTE began in 3GPP in 2005, the framework for LTE was designed to optimize data throughput and voice over internet protocol (VoIP) capacity in macro deployments. Over the course of the subsequent and continuing LTE evolution, additional use cases and scenarios were gradually introduced \cite{G2}, including the support of broadcast services, the extension to heterogeneous networks, the support of device-to-device (D2D) and vehicle-to-everything (V2X) communications, the expansion into unlicensed and shared spectrum, and the addition of machine type communications (MTC) in two flavors targeting different deployments and use cases: enhanced MTC (eMTC) and narrow-band internet of things (NB-IoT).  These additional use cases and scenarios not only enriched LTE's offerings, but also provided a valuable learning experience for future generations of standardization, where the standardization has to consider at the very beginning a variety of different use cases, spectrum types, and deployment scenarios.

\begin{figure}[t]
\centering
	\includegraphics[width=1\linewidth]{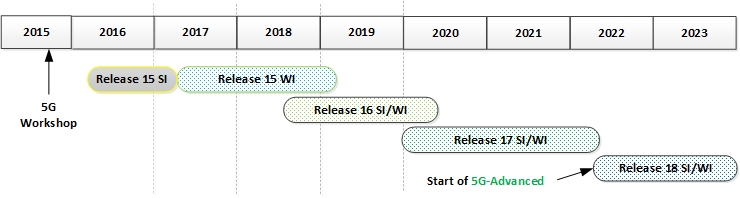}
	\caption{An illustration of 5G timeline in 3GPP.}
	\label{fig:3GPP timeline}
\end{figure}

The first release of 5G, starting with Release 15, introduced a solid baseline for rolling out 5G, first with non-standalone (NSA) networks together with LTE and more recently, standalone (SA) 5G networks with 5G core but without the need of using LTE \cite{I1}. One important aspect of 5G standardization is the easy migration from LTE to 5G in commercial deployments. At the physical layer, 5G NR is designed to be compatible with LTE, e.g., by supporting also a one-millisecond sub-frame duration and the 15 kHz sub-carrier tone-spacing. Fig.~\ref{fig:3GPP timeline} illustrates the timeline for 5G standardization in 3GPP, starting with the first 5G workshop in 2015, which identified three high-level use cases for 5G \cite{Pre5GA1}:
\begin{enumerate}
    \item Enhanced mobile broadband (eMBB),
    \item Massive MTC (mMTC), and
    \item Ultra-reliable and low latency communications (URLLC).
\end{enumerate}

Release 15 was designed to be unified, accommodating and flexible in consideration of the different performance requirements (e.g., in terms of throughput, reliability, latency, coverage, and capacity) of different use cases. It is also scalable, easily adapting to different spectrum and service requirements. In particular, the spectrum supported by 5G is classified into two frequency ranges (FR): FR1 and FR2. FR1 covers 410 MHz to 7.125 GHz while FR2 is further divided into FR2-1 and FR2-2, covering 24.25 GHz to 52.6 GHz and 52.6 GHz to 71 GHz, respectively. 


Releases 16 and 17 addressed the necessary enhancements as a continuation from Release 15 as well as adding capabilities for various vertical segments. The inclusive and future-compatible 5G framework facilitates its evolution in supporting or expanding features enabling new or additional services and use cases, as witnessed in Releases 16 and 17. In particular, these features include \cite{G2}:
\begin{itemize}

\item MTC support, particularly, the accommodation of eMTC/NB-IoT into 5G and the introduction of reduced capability (RedCap) user equipment (UE) targeting new types of devices (e.g., wearables, surveillance cameras, and industrial sensors);
\item Unlicensed and shared spectrum (e.g., 5 GHz and lower 6 GHz bands);
\item Non-terrestrial networks (NTNs) primarily for satellite communications, supporting both eMBB and MTC services;
\item NR sidelink, supporting V2X, public safety, and network controlled interactive services;
\item 5G broadcast and multicast;
\item URLLC and industrial internet of things (IIoT).
\end{itemize}

The solid start of 5G in Release 15 and the expansion of 5G into the above vertical areas in Releases 16 and 17 provide a comprehensive set of standardized features in 5G, equipping operators and vendors with a rich set of enablers for successful commercial deployments. There was a lot of attention on IIoT with work on URLLC as well as on private networks and positioning, enabling the use of 5G with industrial automation. 
At the same time, 3GPP standardization continues its work in boosting the radio performance and improving key metrics such as UE power consumption, uplink coverage or achievable system performance especially with 5G multiple-input multiple-output (MIMO) evolution.

\subsection{5G-Advanced towards 6G}
\label{subsec:5g6g}

3GPP is now entering the second phase of 5G standardization, known as 5G-Advanced, built on the 5G baseline in 3GPP Releases 15, 16, and 17. 5G-Advanced will further expand and extend the 5G capabilities and use cases in many ways, starting with Release 18 to be ready in early 2024 and continuing for further releases in Release 19 and beyond \cite{P20}. The key 5G-Advanced topics are introduced in this paper. 

This paper also looks into future evolution beyond 5G-Advanced, i.e., 6G, and discusses technologies that are expected to come next. Although 6G standardization is not started yet, there have been a plethora of 6G initiatives around the globe, driven by research interest, industry expectations, and strategic government plans. In November 2019, China’s Ministry of Science and Technology set up a working group called “China 6G Wireless Technology Task Force” responsible for the national 6G research and development and another working group consisting of government agencies to promote the development of 6G technology. In the U.S., the alliance for telecommunications industry solutions (ATIS) launched Next G Alliance in October 2020 to advance North American leadership in 6G. Japan established the Beyond 5G Promotion Consortium and Beyond 5G New Business Strategy Center in December 2020 to promote beyond 5G/6G development in Japan. Europe has also launched various 6G initiatives, notably the Hexa-X project launched in January 2021, which aims to shape the European 6G vision and develop key 6G technologies to enable the vision. In June 2021, South Korea established a 6G implementation plan to lay the groundwork for 6G research and development, which aims to push to launch commercial 6G services by around 2028.

This paper is an attempt to summarize and overview many of the exciting developments in 5G-Advanced towards 6G. The rest of this paper is organized as follows. Section \ref{sec:5g-advanced} provides an overview of 5G-Advanced in more detail to explain the Release 18 contents. Further evolution of MIMO is covered in Section \ref{sec:mimo}, followed by positioning evolution in Section \ref{sec:positioning}. Section \ref{sec:topology} discusses the topological evolution with elements like uncrewed aerial vehicles (UAVs), NTNs and network-controlled repeater nodes. Section \ref{sec:xr} describes the support for extended reality (XR) services, which includes augmented reality (AR), virtual reality (VR), or mixed reality (MR) type of services. The sidelink evolution is introduced in Section \ref{sec:sidelink}, while artificial intelligence (AI) and machine learning (ML) for the NR air interface and for the 5G radio access network (RAN) are covered in Section \ref{sec:ai}. The paper continues with the discussion of duplex evolution studies, and  green networks (aiming to reduce the energy consumed by both networks and UEs) in Section \ref{sec:duplex} and Section \ref{sec:green}, respectively. Section \ref{sec:additional} presents additional 6G candidate technologies, followed by the concluding remarks in Section \ref{sec:conclusion}.

\section{5G-ADVANCED: KEY TECHNOLOGIES}
\label{sec:5g-advanced}


The first three 5G releases, as described in Section \ref{subsec:5gpath}, provide a comprehensive package of features in a forward-compatible framework, enabling successful 5G commercial deployments not only for the traditional eMBB services but also expansion into the vertical domains. While it is still imperative to continue evolving 5G in response to the ever-increasing immediate commercial needs, it is also necessary to start exploring new areas to further unleash the potentials of 5G. This is particularly important considering the mounting interests and efforts of 6G in academia and various fora and alliances, as discussed in Section \ref{subsec:5g6g}.

Release 18 will be the first release of 5G-Advanced. Therefore, how to determine the set of features to be included in Release 18 requires meticulous planning and careful discussion. To that end, a 3GPP workshop focusing on RAN-related features was organized in June 2021, which attracted over 500 contributions from about 80 companies, organizations, and research entities  \cite{R18_1}. The workshop was primarily done via conference calls, with more than 1,200 registered participants.

The contributions to the workshop were coarsely classified into three different categories \cite{R18_1}:
\begin{itemize}
    \item eMBB-driven evolution;
    \item Non-eMBB driven evolution;
    \item Cross-functionalities (or new areas) for both eMBB-driven and non-eMBB driven evolution.
\end{itemize}
From these contributions, it can be observed that the proposed evolution directions are generally balanced in terms of:
\begin{itemize}
    \item Mobile broadband evolution vs. further vertical domain expansion;
    \item Immediate vs. longer term commercial needs;
    \item Device evolution vs. network evolution.
\end{itemize}	 
These balanced evolution directions formed the basis for the subsequent discussion, leading to the final approval of the set of features in December 2021 \cite{R18_2}. These features can be roughly categorized into three different categories (eMBB/non-eMBB/new areas), as shown in Table  \ref{tab:RAN_R18_package}. Note that such classification may be subjective and thus arguable. However, it should provide a high-level insight of the overall evolution directions of 5G-Advanced. 
\begin{table}[]
\centering
\caption{\label{tab:RAN_R18_package}3GPP RAN Release-18 package and rough classification}
\begin{tabular}{p{.08\textwidth}p{.36\textwidth}}
\hline \thead{Category} & \thead{Feature} \\
\hline eMBB & -	MIMO evolution (both downlink and uplink) \\
	& - Further NR coverage enhancements \\
	& - Enhancement of NR dynamic spectrum sharing (DSS) \\
	& - NR network-controlled repeaters \\
	& - Multi-carrier enhancements \\
	& - NR mobility enhancements \\
    & - Dual Tx/Rx multi-subscriber identity modules (MUSIM) \\
	& - In-device co-existence (IDC) enhancements for NR and multi-radio dual-connectivity (MR-DC) \\
	& - Mobile terminated small data transmission (SDT) for NR \\
	& - Mobile IAB \\
	& - Further enhancement of data collection for self-organizing networks (SON) and minimization of drive test (MDT) in NR and LTE/NR dual-connectivity (EN-DC) \\
	& - Enhancement on NR quality-of-experience (QoE) management and optimization for diverse services \\
\hline Non-eMBB &	- NR sidelink relay enhancements \\
	& - NR sidelink evolution \\
	& - Expanded and improved NR positioning \\
	& - Further NR RedCap UE complexity reduction \\
	& - Radio enhancements for XR \\
	& - NR NTN enhancements \\
	& - IoT NTN enhancements \\
	& - Enhancements of NR multicast and broadcast services \\
	& - NR support for dedicated spectrum with less than 5 MHz bandwidth for FR1 \\
	& - NR support for UAV \\
\hline New Areas & - Study on AI/ML for NR air interface \\
	& - AI/ML for NG-RAN \\
	& - Study on evolution of NR duplex operation \\
	& - Network energy savings \\
	& - Study on low-power wake-up signal/receiver \\
	& - Study on enhancement for resiliency of gNB-central unit (gNB-CU) \\
\hline  
\end{tabular}
\end{table}

In the sequel, we will dive into the details of some of the Release 18 features, including massive MIMO, positioning, topological aspects, XR, sidelink, AI/ML, duplexing, and green networks. We will introduce these features based on the efforts not only within, but also outside, 3GPP.

\section{MASSIVE MIMO EVOLUTION}
\label{sec:mimo}


In this section, we first introduce the massive MIMO evolution in 3GPP standards and then discuss in detail multi-user (MU-MIMO) related aspects as well as how AI/ML can be used for MIMO operation. After that, we describe future evolution of massive MIMO towards 6G.

\subsection{Evolution of Massive MIMO in 3GPP Standards}
\begin{figure}[t]
\centering
	\includegraphics[width=0.9\linewidth]{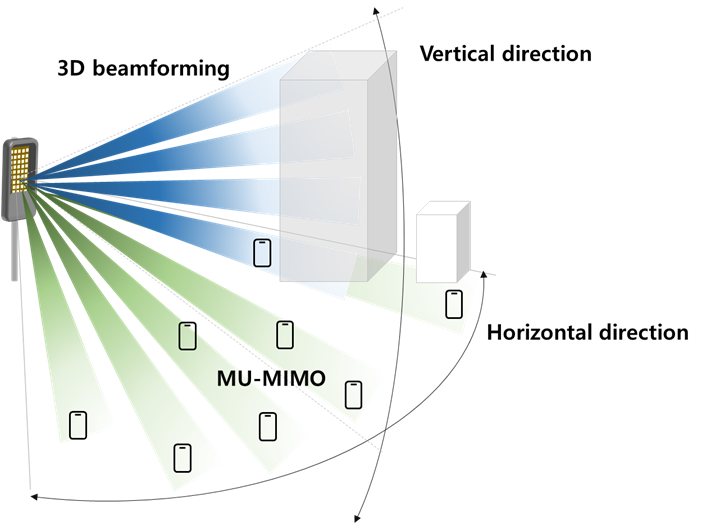}
	\caption{Illustration of the concept of Massive MIMO.}
	\label{fig:Concept of Massive MIMO}
\end{figure}

5G NR was standardized assuming the use of two-dimensional array with a large number of antenna elements, often called massive MIMO, so that beamforming and spatial multiplexing can be performed in a very flexible manner utilizing both horizontal and vertical directions as illustrated in Fig.~\ref{fig:Concept of Massive MIMO}.
This approach was first taken in the full dimension MIMO (FD-MIMO) in LTE-Advanced Pro in Release 13 \cite{M1} and was exploited in the first version of 5G standard in Release 15 \cite{M2}.

By forming a narrow beam, the base station can concentrate its transmission energy along  a desired direction and can improve the strength of the signal received from a desired direction.
Utilizing a two-dimensional antenna array makes it possible to shape beams in both horizontal and vertical directions, resulting in improved capability of distinguishing the signals to and from multiple devices. 
The massive MIMO in 5G is the essential technology for coverage enhancement, improved single user throughput, and cell throughput improvement through the aggressive use of multi-user multiplexing in spatial domain (a.k.a., MU-MIMO).

In 5G, downlink (DL) beamforming and MU-MIMO require the base station to know about the DL channel state, which can be provided from the channel state information (CSI) report from the UE as well as the sounding reference signal (SRS) transmitted by the UE utilizing the channel reciprocity in case of a time division duplex (TDD) system. 
The UE can generate its CSI report by monitoring the CSI reference signal (CSI-RS) transmitted by the base station either per antenna element or per beam. The latter helps reduce the amount of time-frequency resource for CSI-RS transmissions. 
Support of high-resolution CSI feedback plays a crucial role for the MU-MIMO operation. For the purpose of DL data reception, the demodulation reference signal (DMRS) is specified in NR, for which the same precoding as the data is assumed.

For the high frequency band operation, 5G supports analog beamforming in addition to digital precoding and beamforming.
This helps reduce the implementation complexity caused by the use of several hundreds of antenna elements to realize the required coverage by overcoming the increased propagation loss. 
The analog beamforming results in the limitation that all simultaneous channels and signals formed by the same set of antenna elements have to be in the same analog beam.
To cover all directions with this limitation, the so-called beam sweeping operation is supported. Flexible beam management functionality is supported to form and manage beams, which is essential for high frequency (e.g., millimeter wave (mmWave)) band systems.

After the first version of 5G standard in Release 15, 3GPP specified enhancements on various aspects of MIMO functionalities in Release 16 and Release 17 \cite{M3, M4}. 
Some representative examples are described below. 
First, Release 16 specified enhancements for reduction of the CSI feedback overhead via spatial and frequency domain compression.
This makes it possible to improve the cell throughput through MU-MIMO operation without causing too much feedback overhead.
Second, various enhancements for beam management were specified. For example, Release 16 specified layer 1 (L1) signal-to-interference-plus-noise ratio (SINR)-based beam measurement to facilitate interference-aware beam selection.
Measures were specified for reduction of signalling overhead and latency, including L1/layer 2 (L2) signalling-based joint indication of DL and uplink (UL) beams in Release 17, among others.
Third, while Release 15 mainly focused on single TRP-based operation, operation of multiple transmit and receive points (TRPs) was enhanced in Releases 16 and 17.
In particular, Release 16 specified non-coherent joint transmission (NCJT) from multiple TRPs to improve DL data rates and spectral efficiency, while Release 17 introduced the enhancement on CSI report to represent joint channels across multiple TRPs involved in NCJT operation. 
%
Furthermore, Release 17 extended the multi-TRP operation to involve TRPs of multiple cells and specified the support for beam management across multiple TRPs.

The scope of NR MIMO evolution in Release 18 is defined in \cite{M5}. Release 18 gives an emphasis on enhancements for UL MIMO especially targeting non-smartphone type devices such as fixed wireless access (FWA), vehicles, and industrial devices. 
These include the use of 8 transmit antennas to support 4 or more streams per UE, and the simultaneous transmission from multiple panels particularly for mmWave band.
CSI reporting may be enhanced for high and medium UE velocities that were not the main focus of the previous releases.
Continuing the enhancement on multi-TRP operation, the L1/L2-signaling based joint indication of DL and UL beams specified in Release 17 will be extended to the case of multiple TRPs.
CSI acquisition may also be enhanced for coherent joint transmission (CJT), a new target scenario in Release 18, with a number of antennas distributed over multiple TRPs.
To improve the performance of MU-MIMO operation in both DL and UL, a larger number (up to 24) of orthogonal DMRS ports may be specified for cyclic prefix (CP) orthogonal frequency-division multiplexing (OFDM). 

MU-MIMO is a critical operation mode for realizing the performance benefit of the massive MIMO system and was first introduced for LTE-Advanced in 3GPP Release 10 \cite{M30}. Scheduling and precoding of transmitted signals for multiple UEs are fundamental operations of MU-MIMO and can be very challenging in the massive MIMO system due to the underlying massive number of antennas at the base station. Success of AI/ML in various areas motivates the research for its utilization for MU-MIMO scheduling and precoding. In addition, AI/ML is also being investigated to improve the performance of MIMO receiver operations such as channel estimation and symbol detection. The following section introduces the academic research results on MU-MIMO and the application of AI/ML to MIMO.

\subsection{MU-MIMO Operation and AI/ML-Enabled MIMO}
%

%
Unlike other wireless networks, cellular networks need to strike a balance between cell-average throughput and cell-edge throughput. Accordingly, MU-MIMO scheduling becomes a central piece to achieve this objective.
Many signal processing and/or optimization-based scheduling algorithms were introduced in the literature tailoring towards different performance metrics. 
For example, maximum throughput scheduling has been introduced to maximize the sum throughput across all the users in the network \cite{M31} without considering fairness among users.
To address this issue, round robin scheduling has been introduced where users take turns to obtain the radio resources so that the resources can be allocated fairly among all users. 
Since the round robin scheudling does not consider the network throughput, this type of scheduler is also referred to as blind equal throughput (BET) scheduler \cite{M32}.
To cover both network throughput and user fairness, the generalized proportional fair (GPF) scheduler has been introduced \cite{M33}.
For delay-sensitive real-time traffic, the modified largest weighted delay first (MLWDF) scheduler \cite{M34} and the exponential proportional fair (EXP) scheduler \cite{M35} have been introduced for cellular networks. 
In \cite{M36}, a multi-phase optimization-based scheduler for MU-MIMO has been introduced by leveraging large-scale parallel computation to speed up the scheduling speed.

Meanwhile, in cellular networks, the input information to the network scheduler is usually the feedback information from the mobile users. 
Very often this information is either coarse (due to the payload limitation and granularity on channel feedback) or erroneous (due to feedback error). 
Since conventional scheduling methods solely rely on the input information, their scheduling decisions are far from being optimal especially when the input information is not accurate.

Motivated by the success of AI/ML, the scheduling problem can be formulated as an optimal control problem of Markov decision process (MDP), which can be solved using deep reinforcement learning (DRL). 
In \cite{M37}, a DRL-based scheduling strategy has been introduced under the assumption that all resource block groups (RBGs) are allocated to a single user in each transmission time interval.
In \cite{M38}, a DRL-based scheduling strategy has been introduced to adopt the same policy network to allocate every RBG so that the significant training costs and convergence issues are avoided.
In \cite{M39}, a DRL-based scheduler was introduced where the action is restricted to only determine the number of RBGs allocated to each user.
Even though this initial investigation assumes certain idealistic assumptions, they did shed lights on applying ML models to solve the challenging problem of MU-MIMO scheduling in an efficient and resilient way. 
It is expected that the frameworks of DRL and/or multi-agent DRL will play important roles to reduce computational complexity of MU-MIMO scheduling.

MU-MIMO precoding is conducted once the scheduling decision is made. Despite offering better performance, non-linear MU-MIMO precoding schemes, such as dirty paper coding (DPC) or vector perturbation, are not practical for MIMO systems due to their high implementation complexity. 
Simple linear processing techniques have been shown to offer significant performance gains for MU-MIMO systems especially for  massive MIMO systems. 
Maximum ratio transmission (MRT) and zero forcing (ZF)-based methods for MU-MIMO precoding \cite{M40, M41} have been introduced.
Low-complexity MU-MIMO precoding strategies that utilize the direction of arrival (DoA) estimation have also been introduced for massive FD-MIMO systems \cite{M42}. 
Similar to the MU-MIMO scheduling, AI/ML tools have also been utilized in MU-MIMO precoding to alleviate the computational burden especially for massive MIMO systems. 
For example,  \cite{M43} introduces a deep neural network architecture by unfolding a parallel gradient projection algorithm to solve the weighted sum rate maximization problem for a multi-user massive MIMO system.
In \cite{M44}, a graph neural network-based design has been introduced to solve the joint beamforming and antenna selection problem with imperfect CSI for a MU-MIMO system. 
In \cite{M45}, a general form of iterative algorithm-induced deep-unfolding neural network is developed in matrix form to reduce the computational complexity for MU-MIMO precoding.

Other than being utilized in MU-MIMO scheduling and precoding, AI/ML tools have also been recently applied to various other MIMO operations. 
For example, various deep learning methods have been applied to the receiver processing of MIMO operations such as channel estimation and symbol detection. 

\begin{itemize}
    \item \textbf{AI/ML for MIMO channel estimation}: A convolutional neural network (CNN) is adopted in \cite{M46} to learn the parameters of the minimum mean square error (MMSE) channel estimator. 
    Channel estimation was treated as a super-resolution problem in \cite{M47,M48}, where ChannelNet \cite{M47} and ReEsNet \cite{M48} have shown their capability of substantially improving the channel estimation quality. 
    However, these works rely on offline training, where the artificially generated offline training data is assumed to have the same statistical properties as the online testing one, which cannot be guaranteed in a practical system. 
    A reinforcement learning-based successive denoising method is introduced in \cite{M49} to improve the mean square error of least-square channel estimation. This method does not need labelled data for training. However, it requires channel power and number of channel taps as prior knowledge, and the training requires hundreds of OFDM subframes to converge.
    StructNet introduced in \cite{M50} provides a real-time online learning channel estimation method, which only requires over-the-air pilot symbols for training and converges fast. 
    
    \item \textbf{AI/ML for MIMO symbol detection}: A deep learning framework, called DetNET, has been introduced in \cite{M51} for MIMO symbol detection. 
    In \cite{M52}, a deep learning-based symbol detector, MMNet, is constructed through unfolding existing optimization-based symbol detection methods to deep neural networks. 
    A special recurrent neural network (RNN) with extremely low training complexity and overhead, reservoir computing, has been introduced to MIMO-OFDM symbol detection in \cite{M53, M531, M532}.
    Domain knowledge of the MIMO-OFDM waveform has been incorporated in the design of underlying reservoir computing in \cite{M54} to improve system performance without a substantial increase in complexity. 
    Furthermore, it enables a real-time online learning-based symbol detection for MIMO-OFDM systems~\cite{M541, M542}. 
    In \cite{M55}, a multi-mode reservoir computing has been introduced to harness the tensor structure of the massive FD-MIMO channel to boost the detection performance for multi-user massive MIMO networks.
\end{itemize}

\subsection{Future Evolution}
There are active research activities for the evolution of massive MIMO technologies towards 6G. 
This subsection briefly introduces the evolution of massive MIMO for mmWave bands, modular massive MIMO, immense MIMO, and holographic MIMO, among others. 
%

\subsubsection{mmWave massive MIMO} The combination of mmWave and massive MIMO embodies the advantages of both technologies, such as large bandwidth, high beamforming gain, and compact form factor due to small wavelengths \cite{M6,M7,M8}.
For a broadband mmWave signal, the array pattern varies over frequencies, thus a large antenna array is unable to generate beams pointing toward the same direction for a wide range of subcarriers. 
This is known as beam squint \cite{M9, M10}, which poses challenges on transceiver design for mmWave massive MIMO systems.
Three beamforming architectures are generally considered for mmWave massive MIMO: analog beamforming, digital beamforming, and hybrid analog/digital beamforming. 
Analog beamforming, which usually relies on phased arrays, enjoys low hardware cost and low implementation complexity, but it is unable to deal with the beam squint effect since each phase shifter can only provide an identical phase shift over different frequencies.
Advanced codebook design has been proposed to deal with this problem \cite{M9}, but its benefit is limited when either the transmission bandwidth or the number of antennas becomes large.
True-time-delay lines can compensate the delays of different antennas and naturally produce frequency-dependent phase shift, but such a configuration increases hardware complexity and limits the transmission rate.
%
While beam squint can be well handled by digital beamforming, the resultant hardware cost and complexity are high.
Low-resolution analog-to-digital converters (ADCs) can significantly reduce cost and energy consumption, but it becomes difficult to correctly decode the received symbols via spatial oversampling, thus decoding with the beam squint effect under low-resolution ADCs remains an open problem. 
For hybrid beamforming, despite various hybrid array architectures, each radio frequency (RF) chain is connected to multiple antenna elements, thus the corresponding drawbacks of analog beamforming still exist and necessitate further study.

\subsubsection{Modular massive MIMO} Besides the high-frequency spectrum such as mmWave and THz bands, the low band realm is still expected to play an essential role in future generations of wireless communications to offer wide and reliable coverage.
It can be difficult to exploit massive MIMO in the low frequency band due to the half wavelength spacing requirement between antenna elements. To address this issue, architectures like distributed massive MIMO and cell-free massive MIMO \cite{M11,M12}, coordinated multipoint (also known as network MIMO), and distributed antenna systems have been studied. 
Recently, a concept named modular massive MIMO (mmMIMO) has been propounded \cite{M13}, which consists of a variety of predefined basic antenna modules that can be flexibly combined to build a single system. 
Initial simulation results and field tests have shown the huge potential of mmMIMO in improving throughput. 
Nevertheless, a few key issues need to be solved to ensure the commercial success of mmMIMO, including the effect of asynchronous reception among various antenna modules \cite{M14}, DL and UL reciprocity calibration, CSI acquisition and reporting, among others. 
In addition, enormous work is needed to accurately quantify the benefits of mmMIMO. 

\subsubsection{Immense MIMO} 

There is increasing interest in utilizing arrays with extremely large number (e.g., hundreds or even thousands) of antenna elements, exhibited in different flavors including further evolution of 5G massive MIMO (denoted \textit{extreme MIMO} in \cite{ITU-R_FTT}) and extremely large aperture array (ELAA). For convenience, in this article, we group them together under the umbrella term \textit{immense MIMO}. Extreme MIMO is mainly targeted forcentimeter-wave (cmWave), mmWave, and (sub-)THz range. For example, in case of a cmWave range such as 7-24 GHz, extreme MIMO aims to improve the throughout performance while reusing 5G base station sites \cite{M58}. The increased number of antenna elements poses technical challenges including computational complexity, CSI feedback overhead, power consumption, etc. 
%
In ELAA, better spatial resolution is achieved thanks to the increased array aperture\cite{M57,M15}.
Furthermore, the super spatial resolution is likely to make the channels nearly orthogonal among different users, providing a favorable propagation condition. 
Besides enhancing mobile broadband services, the large spatial resolution of an ELAA can also be exploited for spatial multiplexing of an unprecedented number of MTC devices. 
The growth of the number of antennas results in an enlarged array size, which gives rise to new phenomena including the near-field effect and spatial non-stationarity. While there exists some work on the characterization of the near-field effect and spatial non-stationarity for ELAA or similar architectures \cite{M16,M17,M60,M61}, more comprehensive studies and measurement campaigns are in need to thoroughly explore the associated channel models and beam/wavefront management strategies.
\subsubsection{Holographic MIMO} Another evolution trend of massive MIMO is holographic MIMO, sometimes also named as large intelligent surface (LIS)  in the literature, in which the element spacing can be tinier than a half wavelength, and the entire array can be regarded as a continuous electromagnetic aperture in its ultimate form \cite{M18,M19,M56}. 
A continuous aperture is able to create and detect electromagnetic waves with arbitrary spatial frequencies (i.e., wavenumbers) without undesired sidelobes, and offer fine controllability of the array radiation pattern. 
It can also facilitate the sensing and imaging of RF signals \cite{M22}, which is promising in positioning along with integrated sensing and communication \cite{M23}. 
Furthermore, holographic MIMO enables highly efficient near-field power transfer \cite{M18}.
Currently, holographic MIMO can be implemented by either a tightly coupled array of discrete active antennas \cite{M22}, or low-profile metamaterial/metasurface elements that do not require bulky electronic components \cite{M24,M25,M26} and are excited by an RF source.
The exploitation of the full potentials of holographic MIMO requires accurate understanding of its physical properties such as channel characteristics and beamforming methodologies. 
While research work exists on some perspectives of the channel modeling for holographic MIMO \cite{M19, M20, M27}, more comprehensive channel models and real-world measurements are in need to facilitate system design and performance evaluation.
Additionally, the theoretical performance limits of holographic MIMO are worth investigating, which may necessitate new communication theories, such as electromagnetic information theory, i.e.,  the combination of information and electromagnetic theories \cite{M28, M29}.

%

\section{POSITIONING EVOLUTION}
\label{sec:positioning}


In this section, we first describe 5G positioning service requirements and positioning architecture, then provide an overview of the 3GPP Release 17 NR positioning methods and ongoing Release 18 NR positioning work, followed by future outlook on positioning evolution.

\subsection{5G Positioning Services}
5G positioning services aim to support verticals and applications with high positioning accuracies \cite{P1}. 
Example verticals and applications that benefit from high accuracy positioning include eHealth, factories, transportation, logistics, mission critical services, and regulatory requirements.
5G system can satisfy different levels of services and requirements (e.g., decimeter-level positioning accuracy, 99.9\% positioning service availability, and sub-second positioning service latency \cite{P2}).

5G system supports the combination of 3GPP and non-3GPP positioning technologies such as global navigation satellite system (GNSS), terrestrial beacon system, sensors, and wireless local area network (WLAN) and Bluetooth-based positioning. 
The service-based architecture of 5G core network used for positioning services and corresponding network functions (NFs), NF services, and procedures are specified in \cite{P3}, while the UE positioning architecture, functional entities, and operation to support positioning methods in next generation RAN (NG-RAN) are defined in \cite{P4}. 


\begin{figure}[t]
\centering
	\includegraphics[width=0.9\linewidth]{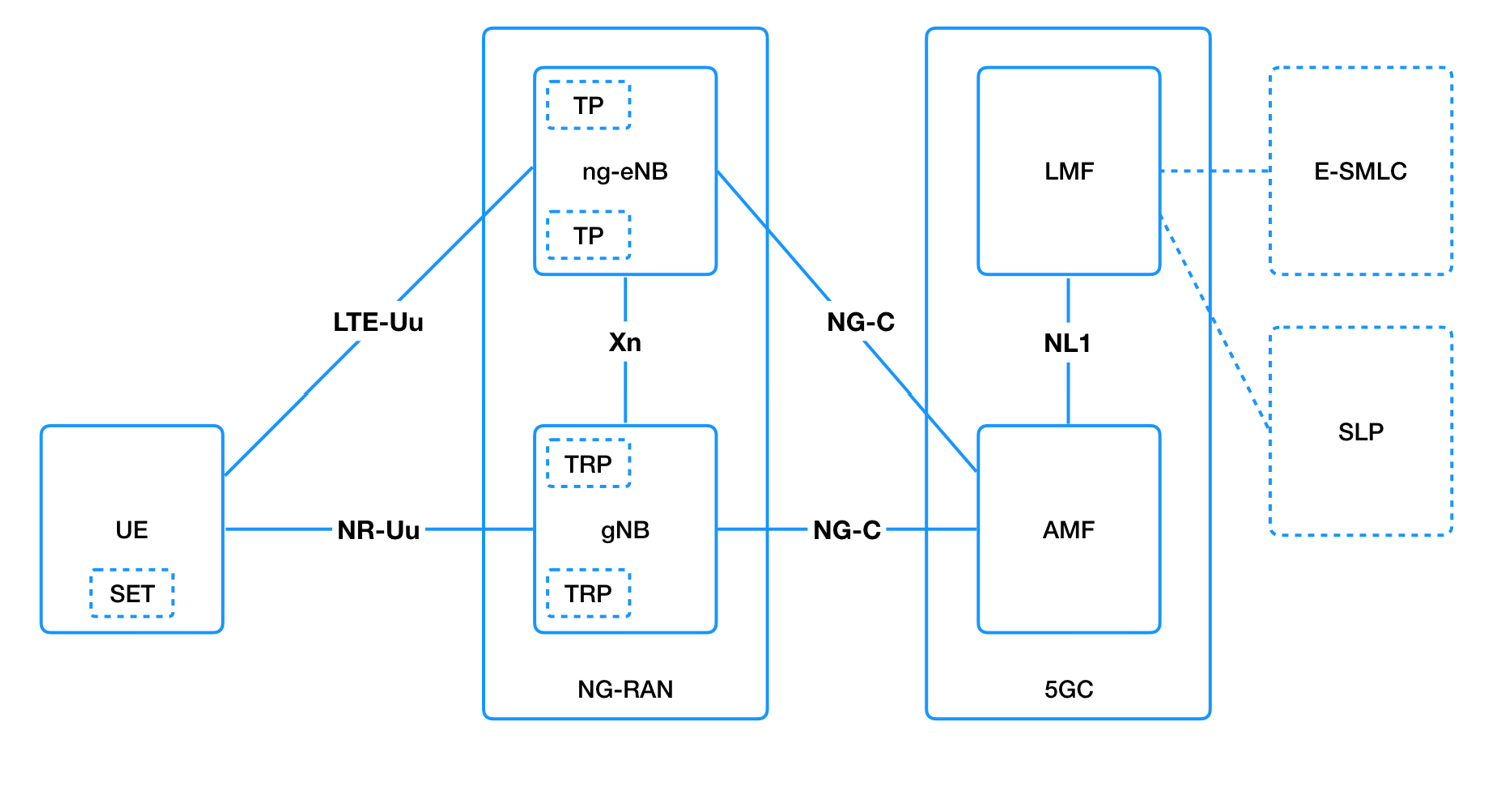}
	\caption{UE positioning architecture applicable to NG-RAN.}
	\label{fig:UE positioning architecture applicable to NG-RAN}
\end{figure}
Fig.~\ref{fig:UE positioning architecture applicable to NG-RAN} shows the UE positioning architecture applicable to NG-RAN \cite{P4}. The positioning information may be requested by and reported to a client within or attached to the core network or in the UE.
When an access and mobility management function (AMF) initiates or receives a request for positioning service for a target UE, the AMF sends a positioning service request to a location management function (LMF) via the NL1 interface.
The LMF processes the position service request by, e.g., transferring assistance data to the target UE and position estimation of the target UE.
An LMF may have a proprietary signaling connection to an evolved serving mobile location center (E-SMLC) which may enable the LMF to access information from evolved universal terrestrial radio access (E-UTRA).
An LMF may also have a proprietary signaling connection to a secure user plane location (SUPL) location platform (SLP) which is responsible for positioning over user plane. 
A UE supporting SUPL is known as SUPL enabled terminal (SET). The NG control plane interface (NG-C) connects the AMF to an NG-RAN node (either a gNodeB (gNB) or a next-generation eNodeB (ng-eNB)).
The connection between the gNB and the ng-eNB is via the Xn interface.
The target UE is connected to the ng-eNB via the LTE-Uu interface and to the gNB via the NR-Uu interface.
The TRP in NG-RAN can be a transmission point for transmitting downlink positioning reference signal (PRS), a reception point for performing uplink SRS measurements, or a combination of both a transmission point and a reception point.
The LTE positioning protocol (LPP), initially defined for LTE, has been extended to support the positioning signaling between an LMF and a target UE in 5G positioning \cite{P5, P23}.
The positioning signaling between an LMF and an NG-RAN node (gNB or ng-eNB) is transported by the NR positioning protocol-annex \cite{P6}.

\subsection{NR Positioning}
Determining the position of a target UE involves two main steps: measurements and position estimation based on the measurements.
The position estimation is usually computed with respect to network nodes with known positions.
In UE-assisted mode, the UE provides the measurements to a location server which computes the position estimation. 
In UE-based mode, the UE computes the position estimation based on the measurements. 
In network-based mode, the network performs the measurements and computes the position estimation.

Positioning errors occur because the measurements are affected by noise, interference, multipath effects, etc.
NR provides improved positioning accuracies by supporting larger signal bandwidths, denser network deployment, and more antennas for transmission and reception \cite{P7}.
Larger signal bandwidth improves signal temporal resolution. Denser network deployment increases the probability of line-of-sight (LOS) channel conditions between TRP and UE.
More antennas for transmission and reception improve signal directivity and thus facilitate angular measurement based positioning methods.

The main NR positioning methods are summarized as follows \cite{P3}.
\begin{itemize}
    \item \textbf{Enhanced cell ID (E-CID)}: In cell ID based positioning, the position of the target UE is estimated to be the position of the UE’s serving base station.
    E-CID enhances the performance of cell ID based positioning by using additional radio resource management (RRM) measurements such as reference signal received power (RSRP) and reference signal received quality (RSRQ).

    \item \textbf{Downlink time difference of arrival (DL-TDOA)} positioning is based on time-of-arrival (TOA) measurements of DL PRSs from multiple TRPs received at the UE. 
    The TDOA values, referred to as DL reference signal time difference (DL-RSTD) measurements, are calculated from the TOA measurements and can be used to compute position estimation.

    \item \textbf{Uplink time difference of arrival (UL-TDOA)} positioning is based on TOA measurements of a UL signal from the UE received at multiple TRPs.
    Since the UL TOA is measured at the TRPs relative to a common time reference, the TOAs are referred to as relative TOA (RTOA) measurements.
    The RTOA measurements of the TRPs are sent to a location server, which calculates TDOAs and uses the TDOAs to compute position estimation.

    \item \textbf{Multi-round trip time (multi-RTT)} positioning is based on two-way TOA measurements, which are used to derive the UE Rx-Tx time difference and gNB Rx-Tx time difference measurements. 
    The RTT between the UE and the gNB can be determined from the reported   UE Rx-Tx time difference and gNB Rx-Tx time difference measurements. 
    Position estimation can be computed by using multiple RTT values.

    \item \textbf{Downlink angle-of-departure (DL-AoD)} positioning is based on per-beam RSRP measurements of downlink PRSs from multiple TRPs received at the UE.
    The per-beam RSRP measurements, combined with additional information such as the TRP positions and DL PRS beam information (e.g., beam azimuth and elevation angular information), can be used to estimate AoD values, which in turn can be used to compute position estimation. 

    \item \textbf{Uplink angle-of-arrival (UL-AoA)} positioning is based on AoA measurements of a UL signal from the UE received at multiple TRPs. 
    The azimuth and zenith of AoA measurements of the TRPs are sent to a location server, which computes position estimation.
\end{itemize}

NR PRSs are fundamental for positioning measurements.
A DL PRS resource can be located anywhere in the frequency grid and has a configurable bandwidth with a comb-like pattern in the frequency domain. 
The DL PRS resource may span multiple consecutive symbols in a slot and have multiple repetitions within one transmission period.
The repetitions of the DL PRS resource are transmitted with the same DL transmit beam which allows the UE to sweep its receive beams over the repetitions of the DL PRS resource.
A TRP can transmit multiple DL PRS resources, each of which may be transmitted with a different DL transmit beam. 
The UL PRS design is based on UL SRS with enhancements for the positioning purpose, such as spatial relations and transmission power control with respect to neighbor TRPs.

In Release 18, 3GPP continues NR positioning evolution \cite{P8}. 
The scope of the work includes studying solutions for sidelink positioning, examining the positioning support for RedCap devices, and investigating solutions to further improve accuracy, integrity, and power efficiency for 5G positioning services. The detailed solutions are now part of Release 18 \cite{S10}.

Positioning is a valuable service that can find applications in diverse use cases. As described in this section, 3GPP has been working on enhancing positioning capabilities of 5G systems over several releases. Future new services and applications will demand ultra-high accuracy positioning (e.g., below centimeter-level accuracy) within a few tens of millisecond latency, which will exceed the positioning capabilities of the current 5G systems. Besides positioning accuracy and latency, other metrics such as reliability, availability, power consumption, scalability, security, and privacy will also be essential design considerations for future positioning services. The following section introduces the ongoing research on positioning, which might become part of the 3GPP positioning work on the path to 6G.

\subsection{Future Evolution}

To meet the diverse extreme positioning requirements on the path to 6G, it is vital to utilize a combination of positioning technologies and exploit different signals such as satellite signals, communication signals, ultrasonic sound, and visible light \cite{P9}. As an advanced technology, large antenna arrays will become more prevalent across different frequencies towards 6G. They can provide high spatial resolution, thus capable of achieving high positioning accuracy. Reconfigurable intelligent surfaces (RISs) that reflect radio waves in preferred directions can also be exploited for localization and mapping \cite{P24}.
LOS/non-LOS (NLOS) path identification in multipath propagation environments is critical for geometry-based positioning methods.
THz technology with ultra-wide signal bandwidth and large antenna arrays can achieve extremely fine time and spatial resolution, enable accurate LOS/NLOS detection, and thus have the potential to improve positioning accuracy significantly \cite{P10}.
In addition, THz sensing can be used to construct 3D images of the environment.
It can enable simultaneous location and mapping (SLAM) by combining high-resolution RF imaging with range and Doppler information \cite{P11}.
THz positioning techniques hold great potential for supporting ultra-high accuracy positioning services on the path to 6G.
Further development of the current SLAM techniques towards exploiting THz localization will be an essential evolution direction for improving positioning services.

Many geometry-based positioning methods rely on TOA or TDOA measurement of the signals.
Tight synchronization down to the sub-nanosecond level is vital for these methods to achieve centimeter-level positioning accuracy \cite{P12}.
Further advancement of synchronization technology will be crucial to support ultra-high accuracy positioning.
Carrier phase positioning (CPP) is another promising positioning technique that uses the phase-locked loop to measure the phase information of received signal and derives the geometric distance between transmitter and receiver from the measured phase. It has been used in GNSS to achieve centimeter-level positioning accuracy \cite{P13}.
However, it is challenging for GNSS to offer centimeter-level positioning accuracy in dense urban or indoor areas. 
Incorporating CPP into cellular positioning was studied in \cite{P8} and CPP is being included together with other positioning enhancements for 5G-Advanced as part of Release 18 \cite{S10}. CPP for cellular positioning will also be an interesting evolution direction towards 6G \cite{P14}.

Traditional geometry-based positioning methods have difficulty in achieving high positioning accuracy in scenarios with heavy NLOS paths, such as indoor factory environments.
AI/ML-based positioning with fingerprinting or ray tracing can overcome the challenge and has attracted much attention \cite{P15} \cite{P16}. 
Indeed, 3GPP studies AI/ML-based positioning as one of the use cases for augmenting the 5G air interface with AI/ML in Release 18 \cite{P17}.
However, to make AI/ML-based positioning a success, several key aspects require further investigation, such as data collection for AI/ML model training, performance validation, and model generalization capability.

Last but not least, though data communication and positioning can coexist in mobile networks, their integration has not been tight thus far.
With the emergence of integrated sensing and communication towards 6G (see Section \ref{subsec:jcs} for more detail), data communication and positioning integration may become tighter in 6G \cite{P18}.
Joint design of communication and positioning to improve both functionalities will be an important evolution direction on the path to 6G and beyond \cite{P19}, which will be further discussed under the category of joint communication and sensing (JCS) in Section XI-B.

\section{TOPOLOGICAL EVOLUTION}
\label{sec:topology}

In this section, we introduce the major topological evolution in 3GPP standards including integrated access and backhaul (IAB), repeaters, UAVs, and NTNs. After that, we describe future topological evolution towards 6G, with a focus on RIS.

\subsection{IAB Evolution}

The support for the IAB node was introduced in Release 16 specifications following the study as captured in \cite{T1}. IAB aims to provide an alternative for fiber or microwave link based backhaul, by using 5G radio interface based either in-band or out-of-band backhauling. Additional enhancements for IAB in Release 17 \cite{T_R17} include improved robustness and load balancing, and reduced service interruption by enhancing topology adaptation, routing, and transport. Duplex enhancements (i.e., simultaneous transmission and/or reception of child and parent links for an IAB node) and efficiency enhancements (e.g., power control enhancements, extension to frequency/spatial domain multiplexing among backhaul and access links) were also introduced.

Release 18 work on IAB \cite{T2} consists of enhancing IAB operation with mobile IAB nodes.
Such nodes would be assumed to be placed in a vehicle (e.g., in a bus) providing coverage for UEs in the vehicle or in the close proximity of the vehicle.
The UEs have no visibility regarding whether the serving cell is a mobile IAB node or not (i.e., vs. a regular gNB), although the service may entail increased latency due to the extra hops in IAB for the connection.

\subsection{Repeater Evolution}
Traditional RF repeaters \cite{T_RFRepeater} are simple but their usage in TDD bands may be compromised, since these repeaters merely boost the received signals irrespective of link directions. A ``smarter'' repeater, on the other hand, can take into account the control information from a gNB, e.g., the TDD configuration in use, such that a transmission may be omitted or appropriately directed as desired, depending on the TDD configuration. These smart repeaters are also known as network-controlled repeaters \cite{T_NCR}.

In Release 18, the support for network-controlled repeater would be done by enabling necessary control information from the network towards the repeater. The necessary information may include the UL/DL TDD configuration, repeater ON/OFF configuration, and beamforming for improved repeater operation\cite{T_NCR}. The work is based on the outcome of the study in 3GPP  as described in \cite{T3}, with the results captured in \cite{T4}.
Controlling the repeater's ON/OFF status is important, e.g., for energy savings when the carrier frequency is turned off for a period of time. There is also a possibility to control the gain and power used by the repeater for more efficient network interference management. 

\subsection{UAV in 5G}
In Release 18, the work on connected UAV covers several key aspects \cite{T5, P22}. 
UAVs are supported to report to networks information such as flight path or height for proper operation. UAV identification is important to help control where the UAVs fly, which can be done by identifying UAVs which could be causing problems (e.g., flying in areas not permitted for UAV use) to ensure UAVs to fly only in authorized air space. 
3GPP is also addressing remote UAV identification broadcast to allow for meeting requirements expected to be part of regulatory requirements in some regions. Besides the pure UAV identification broadcast, considerations have also included potentially more elaborated aspects of detecting other UAVs and avoiding them in order to prevent collisions.

\begin{figure}[t]
\centering
	\includegraphics[width=0.9\linewidth]{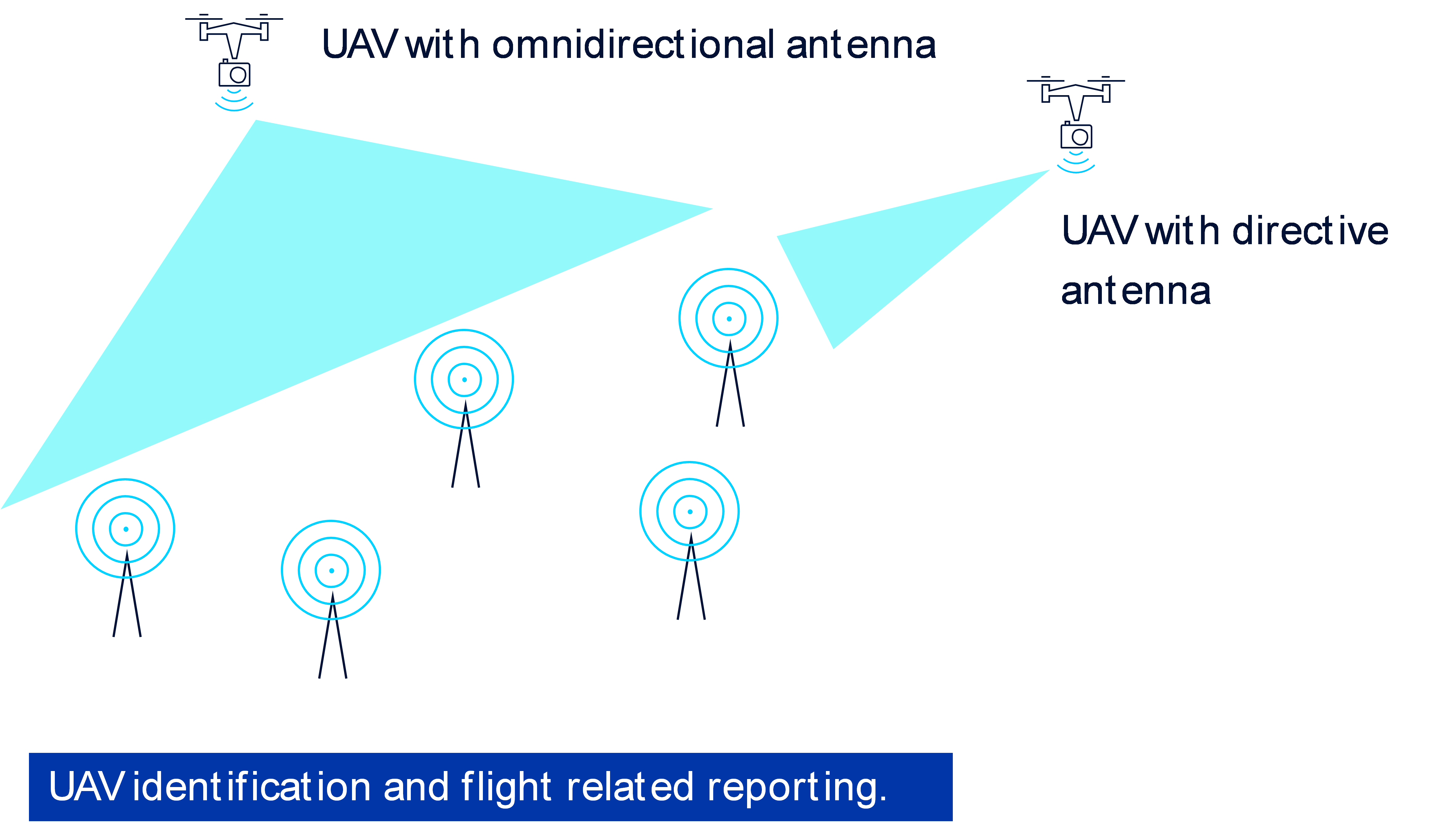}
	\caption{Operation with different UAV antennas.}
	\label{fig:Operation with different UAV antennas}
\end{figure}

It is also critical to consider the impact of UAVs on the network, as UAVs associated with heavy traffic (e.g., video feed) may easily create lots of interference in the network. This is due to the fact that a UAV up in the air may have LOS connections with many cells.
With the use of 5G beamforming capabilities,  directive/beamforming antennas may replace omni-directional transmission such that data flow from a UAV can be more directive, with this aspect being studied also for Release 18. 
This is illustrated in Fig. \ref{fig:Operation with different UAV antennas}.
In such a case, the interference caused to the network can be heavily reduced. In addition, UAV operation can be more reliable since UAVs would experience less interference from different gNB transmitters.  Further aspects of UAV operation have been discussed, including optimization of handover procedures, especially for conditional handover in connection with UAV connectivity. In case that a UAV has a predetermined flight path (and known by the network), one could expect more possibilities to prepare for handover in advance.
%

\subsection{NTN Evolution}

     Besides terrestrial topological evolutions, 3GPP has been working on non-terrestrial topological evolution to support satellite communication. Historically, the terrestrial and satellite communication systems have been developed and evolved independently. Despite its high data rate and low latency, the terrestrial mobile system covers about merely 20\% of the land area which is only 6\% of the entire earth surface \cite{T18}.
     In contrast, the satellite system can offer global coverage and higher survivability in case of disasters such as earthquakes, but suffers long transmission distance hence high latency. 
     Therefore, 5G strives for the integration of terrestrial and satellite communication systems, aiming at obtaining ubiquitous coverage and high quality of service.  

     %
     The most popular conventional satellite communication standard is the digital video broadcasting – satellite – second generation (DVB-S2) standard and its extension (DVB-S2X) from the European Technical Standards Institute (ETSI) \cite{T20}.
     Meanwhile, 5G NR based satellite communication has been widely discussed by standard organizations such as 3GPP and International Telecommunication Union (ITU) \cite{T26}.
     Compared with 5G NR, the target spectral efficiency is lower in DVB-S2X, and certain functions, such as uplink synchronization, hybrid automatic repeat request (HARQ), control channel, discontinuous reception (DRX), RRM measurement, mobility management and core network, are in lack.
     Therefore, DVB-S2X is more suitable for fixed wireless access rather than mobile communication service.

     In 3GPP, NTN was first introduced in 5G NR Release 17 for a unified standard for both the terrestrial communication and the satellite communication, the details of which can be found in \cite{T21,T22,T23}.
     Compared with terrestrial communication, satellite communication has its unique characters in terms of propagation channel, propagation delay, satellite mobility, cell radius, multi-layer coverage, and so on \cite{T25}. The 3GPP NTN work has been addressing the following main technical challenges: 
     \begin{itemize}
         \item \textbf{Challenges in transmission systems}: In the satellite-ground integrated network transmission, Doppler frequency shift, frequency management and interference, power limitation, and timing advance are urgent problems to be solved. 
         For Doppler frequency shift, 5G adopts multi-carrier OFDM in the transmission system, and its subcarrier spacing design does not consider the impact of large Doppler frequency shift, which can bring interference between subcarriers.
         In terms of power limitation, it is necessary to ensure high frequency band utilization and reduce the signal peak-to-average ratio.
         With regard to timing advance, the rapid change of wireless link transmission delay may necessitate dynamic updates of the timing advance of each terminal to ensure that all UL transmissions are synchronized. 
         \item \textbf{Challenges in access and resource management}: Taking LEO satellites as an example, the RTT between an LEO satellite and an earth terminal ranges from 1.2 ms to 13.3 ms, while that for typical terrestrial communications is only up to 1 ms. The long delay of the satellite-ground integrated network brings challenges to the access control, HARQ, and other processes of medium access control (MAC) and radio link control (RLC) layers.
         In terms of access control, in order to support the effective integration of terrestrial and satellite systems, it is necessary to design reasonable access mechanisms such as pre-grant, semi-continuous scheduling, and grant free access.
         For HARQ, which has strict requirements on time, the round-trip time usually exceeds the maximum timer length of HARQ.
         In the scheduling process of MAC and RLC layers, the long delay of the satellite system will also affect the timelines of scheduling, thus its scheduling delay parameters need to be adjusted.
         \item \textbf{Challenges in mobility management}: In the satellite-ground integrated network, the challenge of mobility management is even severer.
         According to the communication level, it can be divided into network-level handover and link-level handover.
         According to the application scenarios, it can be divided into inter-cell handover on the ground, handover between satellites and ground cells, handover between satellite cells, and inter-satellite handover. 

     \end{itemize}

The main solution standardized in the 3GPP Release 17 to address the above challenges is to enable the NTN network to broadcast satellite ephemeris-related data. Furthermore, the NTN UE is assumed to be equipped with GNSS capabilities. With ephemeris data and GNSS capabilities, the NTN UE can calculate the relative speed and RTT between the UE and the satellite to precompensate its UL frequency and transmission timing. UE can also utilize the ephemeris data to predict the trajectory of the LEO satellites over time, facilitating mobility management in NTN networks. We refer interested readers to \cite{T26} for more details.

The latest Release-18 NTN objectives focus on the applicability of the solutions developed by general NR coverage enhancement to NTN, particularly for uplink channels \cite{T_NTNR18}. Additional evolution of NTN in 3GPP is possible to further user experience. In particular, at certain point in the future, satellite communication may be integrated within 6G systems, deployed at high, medium and low orbits, and working jointly with terrestrial communication. The terrestrial network may be the core of the integrated system and controls the entire space-based communication system.

\subsection{Future Evolution through RIS}
\begin{figure}[t]
\centering
	\includegraphics[width=0.9\linewidth]{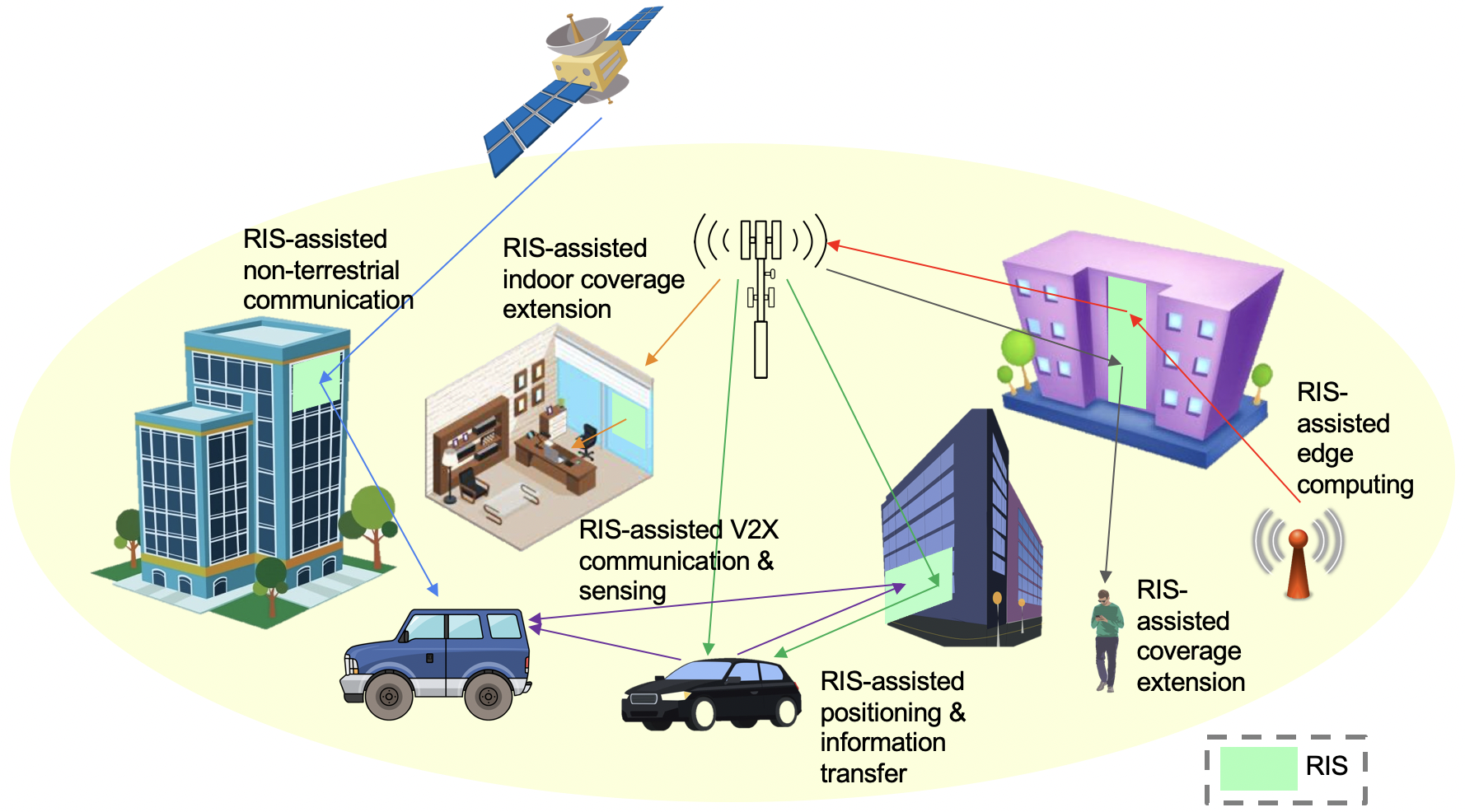}
	\caption{Illustration of anticipated typical use cases of RISs.}
	\label{fig:Typical use cases for reflective and refractive RISs}
\end{figure}

    We anticipate that 3GPP will continue topological evolution to embrace emerging technology trends on the path to 6G. One potential major topological evolution can be the inclusion of RIS. Broadly speaking, RIS is an umbrella term for two-dimensional metamaterial-based arrays that can manipulate electromagnetic waves via anomalous reflection, refraction, polarization transformation, among other functionalities ~\cite{T6,T7}.
   Herein we focus on RISs configured as anomalous reflective and/or refractive surfaces that can customize the propagation environment by reflecting and/or refracting signals to desired directions. 
     Fig.~\ref{fig:Typical use cases for reflective and refractive RISs} illustrates typical use cases for RISs. 

    Among the various types of RISs, reflective RIS is most common, which reflects signals from the source node to the destination node while consuming little energy, leading to high spectral efficiency and energy efficiency.
     The power allocation, phase shift design, beamforming strategies, channel estimation schemes, and a variety of other aspects regarding reflective RISs have been widely studied in the literature \cite{T8,T9,T10,T11,T12}.
     Note that both reflective RIS and relay can provide range extension. Compared with the conventional relay, RIS has some distinguishing features such as hardware complexity, power budget, noise, and average end-to-end SNR \cite{T13} as listed in 
     Table \ref{tab:comparison_ris_relays}. 
     \begin{table}[]
\centering
\caption{\label{tab:comparison_ris_relays}Comparisons between RISs and relays}
\begin{tabular}{p{.07\textwidth}p{.175\textwidth}p{.175\textwidth}}
\hline \thead{ } & \thead{Relay} & \thead{RIS} \\
\hline Hardware complexity & A series of active electronic components & Low-power and low-complexity electronic circuits (switches or varactors) \\
\hline Power budget & A large amount of active electronic components with high power consumption. \newline
Total RF power is often allocated between the transmitter and the relay. & Nearly passive implementation with low-power active components (switches or varactors).\newline 
Almost all RF power is allocated to the transmitter instead of between the transmitter and the RIS. \\
\hline Noise & Additive noise or loop-back self-interference & Phase noise \\ 
\hline Average end-to-end SNR & Proportional to the inverse of transmission distance & Electrically large RIS: Proportional to the inverse of transmission distance. \newline
Electrically small RIS: Proportional to the inverse of the square of transmission distance \cite{T13} \\
\hline  
\end{tabular}
\end{table}

     Qualitatively, RIS can be regarded as a full-duplex MIMO relay without self-interference and signal power amplification \cite{T14}.
     Overall, RIS-assisted transmission may outperform relay-aided transmission in terms of data rate if the size of the RIS is sufficiently large \cite{T13}, but the performance gain may be limited by the quantization of the phase shifts. Thus, it is still subject to more comprehensive studies to determine whether RISs or relays/repeaters are more suitable for commercial deployment, especially when considering various factors including cost, use cases, and maintenance.
     To strive for better performance and to take advantage of both techniques, some researchers proposed combining relays with RISs and named this architecture as hybrid relay-RIS (HR-RIS), in which a single or a few elements of RIS act as active relays, while the remaining only reflect the incident signals.
     HR-RIS potentially excels at spectral efficiency and energy efficiency in harsh scenarios such as when the transmit power is low and/or when the HR-RIS is located far away from the transmitter \cite{T15}. 

     Another type of RIS deployment is to further consider its transmissive properties to extend its application scenarios \cite{T16,T17}, such as simultaneous transmitting and reflecting RIS (STAR-RIS) \cite{T27}, where part of the incident signal is reflected into the space in front of the RIS and the remaining part is refracted into the other side of the RIS.
     %
     The substrate of an STAR-RIS can be optically transparent such that it does not interfere aesthetically or physically with the surrounding environment or people's line of sight. 
     The power ratio of the reflected and penetrated signals is determined by the hardware structure, and needs to be appropriately optimized to enhance the performance over all users. 

     Besides theoretical analysis and simulations, early prototyping and testing have also been conducted to demonstrate the performance gain brought by RISs, including a recent multi-frequency field trial using off-the-shelf 5G UEs and reference signals defined in 3GPP 5G standards \cite{T28} which showed 15 to 35 dB RSRP enhancement by RISs.

     Despite the great potential of RIS and some pioneering field trials, a number of key challenges need to be handled in practical implementation, e.g.:
     \begin{itemize}
         \item Inconsistent phase offset of an RIS across subcarriers for broadband communications,
         \item Channel estimation and feedback overhead for passive RISs,
         \item Efficient power supply for RISs,
         \item Performance degradation owing to element failure, and
         \item Ratio of active and passive elements in an RIS considering the tradeoff among channel estimation complexity, spectral efficiency, and energy efficiency.
     \end{itemize}
Therefore, compared with its counterparts such as small cells and relays, whether RIS can achieve the eventual commercial success requires further investigation and more extensive trials.

\section{XR EVOLUTION}
\label{sec:xr}

In this section, we will start by describing XR-specific service requirements, followed by several key considerations in serving XR in 3GPP. We will also provide a brief overview of the efforts related to XR outside 3GPP.

\subsection{XR-specific Service Requirements}
XR is an umbrella term covering applications including AR, VR, or other forms of expanded and immersive reality applications. 
The XR as a service  differs from the traditional mobile broadband services, as it demands low latency and high data rates in line with the XR-codec periodicity. In particular, the video codecs have often variable frame rate, which is not necessarily evenly aligned with the frame structures used in 5G networks. A typical frame rate of 60 frames per second (fps) would result into a periodicity of approximately 16.67 ms, which naturally does not map well on the 10 ms 5G frame structure with the transmission time internal of either 1 or 0.5 ms  depending on the frequency band and the resulting sub-carrier spacing in use for the sub-6 GHz frequencies. 

Besides the actual video transmission, it is necessary to enable relatively frequent control or pose update signaling transmission, and to keep the experienced delay below 10 ms, given that in some of the traffic models in use \cite{XR1} the control/pose update signalling is  generated every 4 ms. Furthermore, the XR applications will likely have different detailed timing characteristics, and thus the envisioned solutions need to be configurable enough to match them.

\subsection{Key Considerations for  the Support of XR in 3GPP}
\begin{figure}[h]
\centering
	\includegraphics[width=0.9\linewidth]{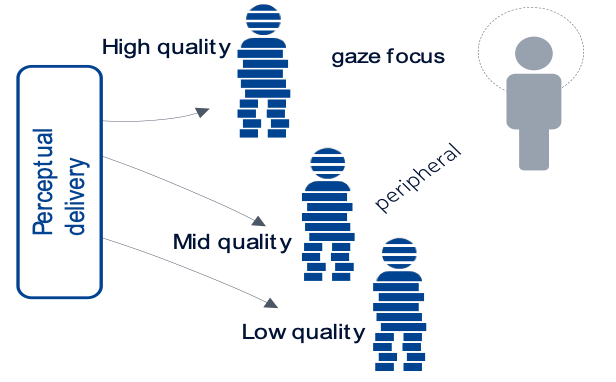}
	\caption{Illustration of XR service.}
	\label{fig:delivery}
\end{figure}

The first key step in supporting XR traffic efficiently is to identify such traffic in the first place, which may be performed for example by extending the 5G quality-of-service (QoS) flow identifiers to enable the detection of the packets that are part of an XR service. Once XR packets are identified,  the scheduling functionality needs to be adapted to the timing characteristics of a given XR service, such as the already mentioned frame rate as well as the jitter that can be tolerated. Additionally, when dealing with services like VR, there is also a trade-off between latency and the required maximum data rate. When the connection has low enough latency, the system allows sending accurately only content for covering  the field of view, as it can react fast enough for the change of gaze focus. A larger latency, instead,  must be compensated by sending also those parts of the view with high accuracy,  which would not necessarily be in the current field of view otherwise. Of course, this also requires that the servers providing the service are not physically located too far from the mobile user, in order to benefit from taking the gaze focus into account, as shown in Fig.~\ref{fig:delivery}. 

Another important factor in XR applications is UE's power saving, as it is desirable to use small and light devices  for XR services. The use of continuous high data rate suited for low latency  causes the mobile devices to consume too much power. Enabling the use of moments when XR data is not coming for DRX operation and other reductions in the receiver processing pipeline allows the UE  to save power and, in turn, improve the usability of XR services. Additionally,  creating some breaks in the data stream  is beneficial  from the network power consumption point of view as well.

To make the XR a real mobile service, the achievable  mobility performance is also important. While the reliability of the connection  is fundamental, it is also essential to minimize the  interruption duration, as otherwise the interruption in the data flow becomes visible in the end-user video stream or the motion related control feedback experiences too much delay. 

Many other improvements foreseen by 5G-Advanced will be of benefit to XR services provision, especially improved uplink coverage and system capacity. In addition to the radio interface aspects, edge computing-related improvements have a direct relationship with XR, as XR application processing in the network needs to be performed sufficiently close to the end user so as to  reduce total end-to-end latency.

3GPP has now established traffic model and evaluation assumptions for XR evaluation over NR \cite{XR2} and studied the potential improvements for XR delivery over 5G-Advanced \cite{XR3} and now XR improvements are being specified for Release 18 \cite{XR3}.  Importantly, this makes delivery of XR services more efficient and, thus, available to a larger number of simultaneous XR users than in the first-phase 5G networks. As XR requires end-to-end handling, not just in the radio access network but also in what concerns edge computing, a significant deal of work addresses  core network-related improvements, especially looking at QoS evolution and application awareness to meet the requirements of XR  services \cite{XR4}. In addition, related to the provision of XR services, 3GPP is  active in developing standards for appropriate codecs of audio and  video, as they will also help deliver XR services for a larger number of users.

\subsection{XR outside 3GPP}
In recent years, XR has attracted a great deal of  interest from both research and commercialization perspectives. While the XR commercial market size is already in the order of billions of US dollars, further significant market growth is expected in the upcoming years  with the development of advanced commercial products.

XR devices may be tethered by cables or supported wirelessly. The former ones facilitate satisfaction of the extreme requirements, but limit users’ mobility and quality of experience. Wireless-connected XR devices instead can leverage advanced wireless technologies to eliminate cables for more enjoyable users' mobility
 anywhere and any time \cite{XR5, XR6}.

While XR may be provided in LTE networks \cite{XR7}, 5G NR is expected to take the VR/AR experience to the next level \cite{XR8}, thanks to  extremely high throughput (multi-Gbps), ultra-low latency (down to 1 ms), and uniform experience (even at cell edge). The work \cite{XR9} underscores the importance of VR technology as a disruptive user case for 5G, and it presents a list of research avenues (e.g., caching, short-range wireless communications, context information and analytics, computer vision and media, etc.) and scientific challenges (e.g., “Shannon-like” theory to maximize users' immersive experience, quality-rate-latency tradeoff, in-VR vs. in-network computation, scalability, privacy, localization, and tracking accuracy). Furthermore, \cite{XR10} lists a set of technical enablers including mmWave communications, edge computing, proactive computing and caching, multi-connectivity, and multicasting, while \cite{XR11} focuses on the reliability and latency achieved for VR services over a THz cellular network. The study \cite{XR12} covers resource management for a network of VR users over small cell networks, where the problem is formulated as a non-cooperative game from which a distributed algorithm is proposed, which leverages ML to predict VR QoS for improved QoS utility. 

\section{SIDELINK EVOLUTION}
\label{sec:sidelink}

In this section, we first introduce the sidelink evolution in 3GPP standards and then discuss its safety applications. After that, we describe future sidelink evolution towards 6G, with a focus on mesh networking.

\subsection{Sidelink Evolution in 3GPP Standards}
\begin{figure}[h]
\centering
	\includegraphics[width=0.9\linewidth]{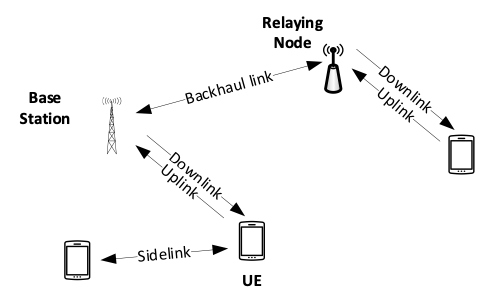}
	\caption{Illustration of various link types in 3GPP.}
	\label{fig:xr_v1}
\end{figure}

Before LTE Release 12, standardization of over-the-air transmissions in 3GPP was primarily focusing on DL and UL Backhaul link in the relaying context was standardized in Release 10 where a relaying node is assumed to be in-band and half-duplex \cite{S1}. With the increasing need of D2D communications, 3GPP started to study and specify D2D in Release 12 \cite{P21} and further evolved it into V2X using the so-called sidelink, as illustrated in Fig.~\ref{fig:xr_v1}.

\begin{figure}[h]
\centering
	\includegraphics[width=0.9\linewidth]{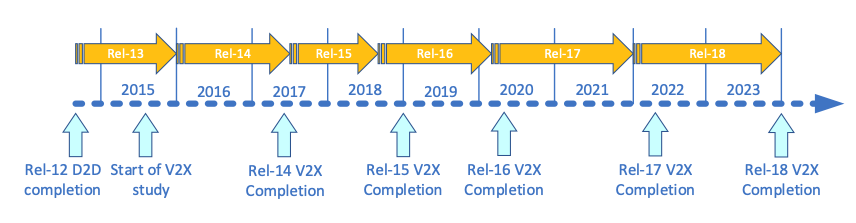}
	\caption{Sidelink standardization timeline in 3GPP.}
	\label{fig:xr_v2}
\end{figure}

Fig.~\ref{fig:xr_v2} illustrates the timeline of sidelink standardization in 3GPP. The standardization started from Release 12 D2D communications, followed by LTE V2X support from Release 14, and 5G NR V2X support from Release 16. The standardization of V2X in LTE and 5G NR is often commonly referred to as cellular V2X or C-V2X.

The D2D standardization in Release 12 covers discovery, synchronization, broadcast and groupcast communications \cite{S2}. Discovery is targeted towards commercial applications, where a discovery signal can be periodically transmitted with roughly 200 bits of information. Communications are for public safety applications only, without any physical layer feedback such as HARQ-acknowledgement (HARQ-ACK) or CSI. The physical layer structure for D2D is completely compatible with LTE for smooth co-existence, where D2D may only occupy LTE UL subframes. The resources for D2D can be either allocated by an eNB, or randomly selected. D2D communications can be supported also for cases when a UE is out of network coverage or partially in coverage. 

Enhanced D2D operations were specified in LTE Release 13 for both discovery and communications \cite{S3}. Discovery was expanded to accommodate inter-frequency and inter-public land mobile network (PLMN) cases, and can be performed with a transmission and reception gap. Out-of-coverage discovery is supported as well. Enhancements to D2D communications include layer-3 UE-based relay, priority handling, and multiple destination transmissions.

Starting from LTE Release 14, V2X is supported in 3GPP. The D2D interface (also known as the PC5 interface) is not only used to support vehicle-to-vehicle (V2V), but also for vehicle-to-infrastructure (V2I), vehicle-to-pedestrian (V2P) and vehicle-to-network (V2N) for beyond safety and collision avoidance \cite{S4}. The primary target spectrum is the unlicensed intelligent transportation systems (ITS) spectrum \cite{S5}.  During the standardization, the maximum vehicle speed was assumed to be 250 km/h, implying a maximum relative speed of 500 km/h. At a carrier frequency around 6 GHz, it would result in a Doppler shift of about 2.7 kHz  and consequently channel variations within a 1 ms subframe, bringing various design challenges.  Wireless resource for V2X can be either scheduled by an eNB or selected in an autonomous manner with the help of sensing. The sensing is performed based on a combination of priority information, energy sensing, and sidelink control channel decoding. Congestion control is also supported such that resource utilization becomes restricted upon congestion based on a channel busy ratio (CBR). Prioritization between the access link transmissions and V2X transmissions may be necessary. Additional enhancements for V2X were also done in LTE in Release 15 \cite{S6}, focusing on aggregating multiple carriers and enhancing autonomous resource allocation. 

The work on 5G NR V2X started in Release 16 \cite{S7}. 5G NR V2X is designed to be backward compatible at upper layers so that it can co-exist with LTE V2X (as in Releases 14 and 15). In particular, Release-16 V2X is designed to support Release-14 and Release-15 V2X for safety use cases. Additionally, Release-16 V2X enables new services such as real-time updates, and coordinated driving, especially those requiring high reliability and QoS support. Different from Release-14 and Release-15 V2X where only broadcast messages are supported, Release-16 V2X additionally supports groupcast and unicast. Spectral efficiency is significantly improved, e.g., via higher modulation order (up to 256-quadrature amplitude modulation (256QAM)) and MIMO transmissions (up to two layers). Similar to the access link, flexible numerologies are also supported, e.g., with subcarrier spacing of 15, 30, 60, and 120 kHz. To cope with extremely high mobility, a range of DMRS patterns are supported with different time-domain densities. 

Enhancements in Release-17 NR V2X \cite{S8} are mainly focused on two aspects: power savings and resource allocation. Power savings are achieved via the support of partial sensing and sidelink DRX operations. Resource allocation can be improved significantly when two or more UEs coordinate with each other and share the resource availability/non-availability information, where the information may be assessed based on various reasons (e.g., sensing and capability). These resource allocation enhancements may bring improved reliability, reduced latency and reduced UE power consumption.

NR sidelink will be further enhanced in Release 18 in a more expanded manner \cite{S9, S10, S11}, including the following areas:

\begin{itemize}
    \item Sidelink carrier aggregation (CA) operation;
    \item Sidelink in unlicensed spectrum;
    \item Enhanced sidelink operation for FR2 licensed spectrum;
    \item Mechanism(s) for co-channel coexistence for LTE sidelink and NR sidelink;
    \item Sidelink relay enhancements;
    \item Sidelink positioning and ranging.
\end{itemize}
With the above items, it is expected that not only the support of C-V2X can be more comprehensive and effective, but also the support of other services (e.g., network controlled interactive services via sidelink) can  be enabled or enhanced.

\subsection{C-V2X for Safety Applications}
Given the impelling need to reduce car accidents and traffic fatalities involving vulnerable road users (VRUs), such as pedestrians, bicyclists and e-kick scooters, safety applications enabled by mobile communications have become crucial to ITS. Such applications are particularly relevant in the presence of obstacles, such as buildings, that prevent a driver or the sensors aboard an autonomous/automated vehicle from realizing the danger in a timely manner. Most of these applications leverage such messages as the ETSI cooperative awareness messages (CAMs) that can be periodically broadcasted by the vehicle's radio interface and include the vehicle’s position, speed, acceleration, and heading \cite{S12}. Additionally, as discussed in \cite{S13}, safety applications can often benefit from messages generated by road infrastructure or smart city entities (e.g., cameras, and traffic controllers or dynamic map generators located at the edge of the network infrastructure), as they can effectively complement the local view that each single vehicle conveys through the transmission of, e.g., CAMs. Several studies have therefore appeared in the literature proposing solutions for safety applications enabled by C-V2X, where the X stands indeed for everything, including other vehicles or VRUs, the network infrastructure, or an edge server. The recent focus on C-V2X as communication technology for connected cars has also been motivated by two important facts. Firstly, as reported in the previous section, the 3GPP C-V2X standardization efforts aim at meeting increasingly challenging requirements for advanced services. Secondly, there is a clear regulation shift towards C-V2X, confirmed by, e.g., the Federal Communications Commission (FCC) regulations that have changed the use of the 5.9 GHz frequency band reserved for ITS operations from dedicated short range communications (DSRC) to C-V2X. 

A useful taxonomy of the types of V2X communication and the applications that they can support can be found in \cite{S14} and is highlighted in Table \ref{tab:v2x}. Specifically, the table lists the main safety applications enabled by C-V2X and the type of communication they require, ranging from periodic V2V/V2I/V2P messages, to V2V/V2I event-driven messages, V2V bi-directional and multi-hop communication, to V2P/V2N bi-directional communication.

\begin{table}[]
\centering
\caption{\label{tab:v2x}Road safety applications supported by different message types}
\begin{tabular}{p{.14\textwidth}p{.3\textwidth}}
\hline \thead{Message Type} & \thead{Supported Applications} \\
\hline V2V periodic & Cooperative collision warning, intersection movement assistance, slow vehicle warning,
cooperative glare reduction, collective perception  \\
\hline V2V event-driven & Stationary vehicle warning, emergency electronic brake lights, queue/traffic jam ahead warning, road condition warning \\
\hline V2I bi-directional & See-through \\
\hline V2V multi-hop & Coverage extension of ITS messages \\
\hline V2P periodic & Pedestrian collision warning to vehicles, vehicle collision warning to pedestrians  \\
\hline V2I periodic & In-vehicle signage, curve speed warning \\
\hline V2I event-driven & Infrastructure based collision warning, warning of vulnerable road user presence, infrastructure based traffic jam ahead warning, infrastructure based road condition warning, roadwork warning \\
\hline V2N bi-directional & Dynamic map download and update  \\
\hline
\end{tabular}
\end{table}

Relevant examples of solutions appeared in the literature, targeting the support of safety applications through the aforementioned types of messages, include \cite{S15} where V2V periodic messages allow for collective perception sharing based on Release-14 LTE PC5 mode 4, and \cite{S16} which envisions mobile base stations deployed in UAV to improve the connectivity offered to vehicles by terrestrial base stations. The use of V2V event-driven messages, combined with CAM transmissions, has been instead leveraged in \cite{S17} to implement a cooperative lane merging application using the LTE PC5 mode 4 technology. Interestingly, \cite{S18} and \cite{S19} have highlighted the limitations of the configuration of PC5 mode 4 interfaces in Release-14 LTE, and, especially, of the autonomous sidelink resource allocation algorithm to transmit V2V event-driven messages -- an issue that has then been overcome in Release-16 NR PC5 mode 2 \cite{S20}. For example, the sensing window of 1 s used in Release-14 LTE PC5 mode 4 works well for V2V periodic messages, but it is too long for asynchronous traffic, in which case a 100 ms-window turns out to perform much better. Furthermore, it is worth mentioning that the combination of V2V periodic and event-driven messages have been successfully leveraged to also support platooning applications \cite{S21}. As far as V2V bi-directional messages are concerned, they are useful in the see-through application, where a vehicle requests and receives light detection and ranging (LiDAR) or camera data from another vehicle, so it can “see” beyond the visibility range of its radio and sensors. To support the exchange of such bi-directional information, \cite{S22} has investigated the use of mmWave communications, while the results of field tests with pre-5G NR Uu interfaces have been reported in \cite{S23}. Finally, V2V multihop messages have been used mainly to extend the propagation of safety messages in vehicular networks, as in \cite{S24} where vehicle clustering and Release-14 LTE PC5 mode 3 were used. 

Many safety applications for vehicles and VRUs are also supported through   V2I messages, which are periodic and event-driven, and sent from vehicles to the infrastructure and then re-broadcasted by the latter through infrastructure-to-vehicle (I2V) or infrastructure-to-pedestrian (I2P) messages so that they can be disseminated over a larger area and leveraged at each mobile entity as input to, e.g., a collision avoidance algorithm \cite{S25, S26, S27, S28}. I2V messages can also be effectively used for in-vehicle replication of road signaling (e.g., traffic light information) and to support autonomous driving by providing vehicles with information generated by road infrastructure entities or traffic controllers \cite{S29, S30}, as well as for dynamic map download and update \cite{S31}. 

\subsection{Future Evolution}

5G NR-based sidelink enables UE-to-network and UE-to-UE relay to significantly improve network performance for C-V2X and public safety. Looking ahead to 6G, it is possible for sidelink to expand existing C-V2X capabilities, to support new use cases such as aerial networking with flexible network architectures. Furthermore, on top of the single-hop relay functionality that has already been supported by the existing 3GPP sidelink, multi-hop routing/networking will allow multiple UEs and UAVs to connect and create a mesh network under various coverage scenarios for different applications and use cases.

mmWave D2D relay has been introduced in \cite{S32} to improve the coverage and spectral efficiency of a wireless network. With the help of D2D relay, it is possible to form a mesh network with dynamic network topologies and architectures. Meanwhile, it is also possible to create a three-dimensional (3D) mesh network by incorporating UAVs and NTN nodes such as high altitude platform stations (HAPS) and satellites. For example, a swarm UAV network has been introduced in \cite{S33} for substantial throughput enhancement. Recently, there is also a tremendous interest in planning novel LEO satellite systems, called mega-constellations, to form a multi-layer mesh network \cite{S32}. Routing and resource allocation are critical for mesh networks and these problems will become even more challenging in a 3D network with multiple layers. Traditionally, cooperative communication and routing strategies \cite{S35, s36, S37, S38} have been introduced for D2D mesh networks to improve the underlying network performance. Random network coding-based cooperative routing strategies have also been introduced in swarm UAV networks to reduce transmission delay with increased communication reliability \cite{S39}. For more complicated multi-layer mesh networks, it is envisioned that AI/ML could be adopted to adjust routing paths \cite{S40}, to conduct distributed and dynamic resource allocation \cite{S41}, and to optimize handover decisions within layers and among layers of the 3D NTN architecture \cite{S42}.

\section{AI/ML EVOLUTION}
\label{sec:ai}

In this section, we first provide an overview of the 3GPP management system for network slicing and data analytics. Then we introduce the 3GPP work on AI/ML for NG-RAN and  NR air interfaces. After that, we discuss AI-native air interface towards 6G.

\subsection{Management System and Data Analytics}
\begin{figure}[h]
\centering
	\includegraphics[width=0.9\linewidth]{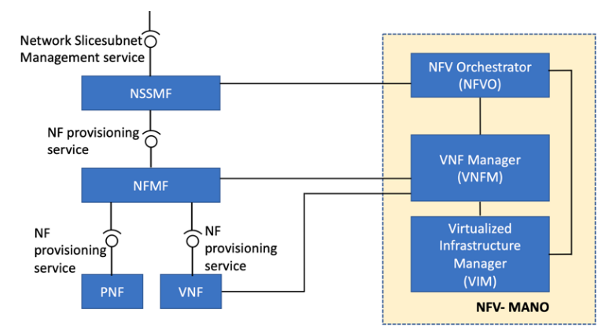}
	\caption{Network slice subnet management with interface to NFV-MANO \cite{A1}.}
	\label{fig:nfv_mano}
\end{figure}

One of the key features of 5G is the ability to support network slicing, i.e., to create multiple logical networks on a shared infrastructure spanning over the core network till the radio access network. Each network slice is composed of one or more NFs, which can be physical or virtual (PNF and VNF, respectively). Furthermore, each network slice is characterized by a set of key performance indicators (KPI) and, possibly, isolation requirements. In order to deploy, orchestrate, and manage a network slice, hence the NFs therein, 3GPP has defined a 5G management system, as illustrated in Fig.~\ref{fig:nfv_mano}. 

A network slice subnet (NSS) provides the management aspects of the NFs managed through the 5G management system along with the corresponding required computing, memory, and networking resources \cite{A1}. The set of NFs, either physical or virtual, composing a network service, represents the actual implementation of an NSS, i.e., a network slice subnet instance (NSSI). The NSS management system in Fig.~\ref{fig:nfv_mano} includes the main components, namely, the NSS management function (NSSMF) and the NF management function (NFMF), as well as the interface with another key element of the 3GPP network functions virtualization (NFV) architecture, i.e., the NFV-management and orchestration (MANO).

\begin{figure}[h]
\centering
	\includegraphics[width=0.9\linewidth]{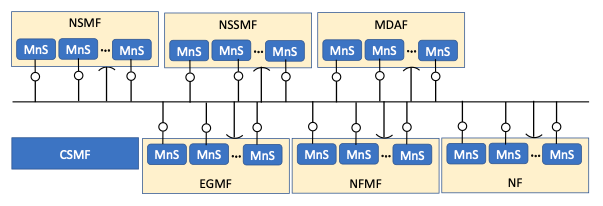}
	\caption{Example of functional management architecture \cite{A1}.}
	\label{fig:func_management}
\end{figure}

Importantly, the 3GPP Release-17 management and orchestration architecture framework \cite{A1} also introduces a service based management architecture (SBMA) for the 5G management system, as depicted in Fig.~\ref{fig:func_management}.  This comprises management functions (MFs) that provide multiple management services (MnS) to manage single functions (NFMF), network slice subnets (NSSMF), communication services (CSMF), and the exposure of such services towards external entities (EGMF). It is worth noticing that the SBMA also comprises the management data analytics function (MDAF), which is in charge of delivering analytics services for automated network management and orchestration. Such data-driven decisions drive the logic of the NSMF and NSSMF, which manage the lifecycle of the network slices and the orchestration of their resources. 

It is also worth to remark that the MDAF receives as input monitoring data collected from the network (gNBs in the RAN or specific core network functions) and its management system on the existing NSS or VNF performance, as well as analytics data offered by the network data analytics function (NWDAF) \cite{A2}. In other words, MDAF and NWDAF operate at the management and at the core network level, respectively, with, e.g., the former being concerned with information on the load level of a slice instance across RAN and core network domains, while the latter with data and control in the core network domain. The MDAF then provides analytics results to drive the decisions related to the NSS life cycle management and the NS orchestration, including RAN and core network domain-related decisions. In particular, the NWDAF not only produces analytics related to the core network domain and network functions and feeds them to the MDAF, but it also leverages the MDA reports to control the core network. Interestingly, a 5G network can feature more than one instance of NWDAF, each associated with a different service area and possibly specialized in providing a certain type of analytics, and with one of the deployed instances acting as an aggregator \cite{A3}. 

\subsection{AI/ML for 5G Radio Access Networks}
Data driven techniques such as AI/ML-based solutions can be a powerful tool to address the challenges of 5G RANs. 5G RANs need to address complex system design and large network optimization problems to meet a wide range of performance requirements, including coverage, data rate, latency, reliability, and user experience. Though a vast number of AI/ML applications in RAN can be up to proprietary implementations and solutions, investigating the needed standards support is essential to accelerate the use of AI/ML techniques in 5G RAN and beyond \cite{A4, A5}.

In Release 17, 3GPP conducted a study to investigate the functional framework for RAN intelligence enabled by further enhancement of data collection \cite{A6}. The study identified a set of high-level principles for AI-enabled RAN intelligence, such as focusing on AI/ML functionality and corresponding types of inputs and outputs while leaving the detailed AI/ML algorithms and models to implementation. Leaving AI/ML algorithms and models to implementation can incentivize vendor competitiveness. The study also introduced a functional framework for RAN intelligence and investigated a set of use cases, including:

\begin{itemize}
    \item \textbf{Network energy saving:} AI/ML algorithms can predict traffic load and energy consumption by leveraging the data collected in RAN. The prediction can be used to help decide cell activation and deactivation strategies to improve network energy efficiency \cite{A7}.
    \item \textbf{Load balancing:} AI/ML algorithms can predict traffic load and automate the optimization of mobility management parameters to improve user experience and system capacity \cite{A8}.
    \item \textbf{Mobility optimization:} AI/ML algorithms can enhance mobility performance by reducing the probability of unintended events (e.g., handover failure, radio link failure). AI/ML algorithms can also predict UE location and mobility trajectory, which can serve as valuable inputs for RRM and traffic steering. 
\end{itemize}

After completing the Release-17 study on AI-enabled RAN, 3GPP continues to conduct the corresponding normative work in Release 18 \cite{A9}. The Release-18 work item aims to specify data collection enhancements and signaling support for AI/ML-based network energy saving, load balancing, and mobility optimization. The enhancements will be introduced within the existing 5G RAN interfaces and architecture. Once the normative part is completed, additional uses cases are expected to be taken under study as part of a new study item for AI-enabled RAN.  

To reap the full benefits of AI/ML for 5G, exploring the potential of augmenting the NR air interface with AI/ML-based features is indispensable. To this end, 3GPP Release 18 includes a study item that investigates AI/ML for NR air interface to improve performance \cite{P17}. The study aims to set up a general framework for AI/ML related enhancements for air interface, such as characterizing the defining stages of AI/ML related algorithms and the life cycle management of AI/ML models, exploring different levels of collaboration between UE and gNB, and investigating needed datasets for AI/ML related enhancements for NR air interface. The study takes a use case-centric approach, focusing on the follow three use cases:
\begin{itemize}
    \item \textbf{CSI feedback enhancements:} AI/ML algorithms can be used for CSI compression in frequency domain and CSI prediction in time domain to reduce overhead and improve accuracy.
    \item \textbf{Beam management:} AI/ML algorithms can be used for beam prediction in spatial and time domains to reduce overhead and latency and improve beam selection accuracy.
    \item \textbf{Positioning accuracy enhancements:} AI/ML algorithms can be used for direct AI/ML positioning (e.g., fingerprinting) and AI/ML-assisted positioning (e.g., the output of the AI/ML model inference is a new measurement or an enhancement of an existing measurement) to improve positioning accuracy.
\end{itemize}

For each use case, the study aims to establish evaluation methodologies and determine KPI to thoroughly evaluate the performance benefits of AI/ML algorithms for the NR air interface. Last but not least, the study assesses the potential specification impact of AI/ML-related enhancements for NR air interface, including physical layer aspects, protocol aspects, and interoperability and testability aspects. Normative work is foreseen for Release 19.

\subsection{AI-native Air Interface}
Moving over 3GPP Release-18 5G-Advanced towards 6G and beyond, it is important to consider AI-native air interface where we will rely on AI/ML models and tools to design individual network components or a purely end-to-end communication system. This paradigm shift in the approach of network design represents a potential disruption to the established cellular systems design procedure, extending to the deployment and operation stages as well. 

\subsubsection{AI-native end-to-end design for the physical layer}
The receiver block in the physical layer is the focus of most research efforts in infusing AI/ML-based solutions to overcome issues such as model mismatch and complexity in traditional model-based processing \cite{A4}. For example, AI/ML-based techniques have been used to efficiently combat the non-linear distortion created by power amplifier \cite{A11} and low-resolution ADC \cite{A12}. They have also been applied to transmitter blocks with the intent of optimizing performance compared to and in addition to existing closed-loop feedback methods. For example, \cite{A13} uses deep learning for combined channel estimation and hybrid precoder design. The AI approach, however, requires that ML methods are used fundamentally to design the entire chain which includes the transmitter, channel and the receiver. There exists work in end-to-end physical layer design which aims to jointly train the transmitter and the receiver, while catering to simpler additive white Gaussian noise (AWGN) \cite{A14} or Rayleigh block fading channel \cite{A15} scenarios. Built on this, \cite{A16} has developed a jointly trained neural network based transmitter and receiver that achieves the state-of-art bit error rate without needing pilot transmissions in practical wireless channel models. Although promising, the AI/ML-based receiver still needs to be trained offline for a significantly large number of frames, which may be an impediment to practical adoption since an online-based operation is critical for cellular networks especially when the adaptation is on-the-fly. 

\subsubsection{AI-native design for the MAC layer}
Given the complex nature of tasks that the MAC layer is responsible for including, but not limited to, user selection, random access, resource allocation, etc., the solutions to these are mostly heuristic in nature. Since it is extremely challenging to identify optimal solutions for these problems in real-time, there is a large potential to leverage AI/ML-based solutions in addressing MAC layer problems. In this context, reinforcement learning usually finds itself best placed as the ideal tool which can handle the time-varying network conditions that may be attributable to the users or the channel. There are examples of ML-based solutions applied to a wide gamut of MAC layer operations such as reinforcement learning-based approaches for random access \cite{A17} and for spectrum access/sharing \cite{A18}, federated learning for minimizing breaks in presence in wireless VR \cite{A19}, and predictive resource allocation in IoT networks \cite{A20}. However, it is fair to say that a truly “AI-native” MAC design that learns and evolves over time still needs significant work. In some ways, this involves learning an entire protocol from scratch and would need to utilize tools from deep multi-agent reinforcement learning \cite{A21}. Furthermore, sample efficiency and convergence rate of the corresponding reinforcement learning tools need to be significantly improved for them to be relevant for practical networks for 6G and beyond to realize AI-native design.

\section{DUPLEX EVOLUTION}
\label{sec:duplex}
In this section, we will first provide an overview of duplex evolution in 3GPP, starting from the standardization of traditional frequency division duplex (FDD) and semi-static TDD operations, followed by dynamic operations for better traffic adaption, and the most recent study on simultaneous transmission and reception at a gNB. We will then explore the related research outside 3GPP.

\subsection{Duplex Evolution in 3GPP Standards}
\subsubsection{FDD vs. TDD}
FDD and TDD have been the two duplex technologies adopted in 3G, 4G, and 5G standards of 3GPP. In FDD, a paired spectrum is allocated for DL, where a base station transmits and a device receives signals, and UL, where a device transmits and a base station receives signals. TDD does not require a paired spectrum and partitions the spectrum into DL and UL in the time domain.

It would be natural to consider FDD for the case of symmetric traffic between DL and UL, e.g., when voice calls dominate the wireless traffic.
Furthermore, the simultaneous existence of DL and UL using a paired spectrum enables simple and efficient system operations. 
An example is that a device can send to its base station an acknowledgement of success or failure of DL data reception while keeping a fixed time interval between DL reception and UL transmission as specified in 3G high speed downlink packet access (HSDPA), 4G LTE, and 5G NR standards \cite{D1, D2, D3}. 

TDD has the benefit of offering flexible deployments with an unpaired spectrum and hence becomes more attractive when it is difficult to find a pair of frequency bands. 
This is a major reason for choosing TDD to deploy 5G systems using a new spectrum such as 3.5 GHz or 28 GHz bands where it is difficult to find a paired spectrum. 
Furthermore, the channel reciprocity required for operating massive MIMO without excessive feedback overhead is an important advocate for adopting TDD.

Provisioning of different configurations of DL and UL resources in time domain depending on the traffic demand of each link has been understood as an important benefit of TDD. In practical TDD cellular systems with macro cell deployments, on the contrary, it has been typical that a pre-defined DL-UL resource partitioning in time domain and the same transmission direction (UL or DL) are maintained in all cells of the network of a mobile network operator (MNO) to prevent the cross-link interference between DL and UL of neighboring cells. A base station’s DL transmission could cause serious interference to UL reception of a neighboring base station. The UL transmission from a UE of a cell could become a strong interference to the DL reception of another UE of a neighboring cell especially when they are located closely in a cell boundary area. 
Furthermore, MNOs using adjacent bands typically align their DL-UL configurations to prevent interference between their networks. 
With the increasing amount of mobile traffic generated in hotspots and indoor environments, there is increasing interest in  heterogeneous networks consisting of macro and small cells. In such scenario, it could become more useful to have dynamic adaptation of DL and UL allocation in TDD operation of a small cell depending on its own traffic situation independently.

\subsubsection{LTE enhanced interference mitigation and traffic adaptation (eIMTA)}
3GPP made standards enabling the adaptation of configuration of DL and UL subframes per 10 ms frame duration in LTE Release 12, which was known as eIMTA \cite{D4}. Additional measures for handling of the cross link interference (CLI) between DL and UL were specified in LTE standards \cite{D2}. The eIMTA operation could cause severe variations in SINR between consecutive transmission time intervals even though the radio channel does not change. To have proper awareness of SINR fluctuation due to eIMTA, separate CSI measurements for the DL subframe, and the flexible subframe that can be either DL or UL were introduced. In order to avoid the severe performance loss due to the CLI, UL power control was enhanced to allow having different transmit power levels between the fixed UL subframe and the flexible subframe.

\subsubsection{NR dynamic TDD}
As a further evolution of TDD operation in 5G, 3GPP specified dynamic TDD feature in Release 15 NR standard, where L1 signaling of slot format indication designates each symbol of a slot to be DL (‘D’) symbol, UL (‘U’) symbol, or flexible (‘F’) symbol according to Table 11.1.1-1 in \cite{D3}. ‘F’ symbol could be either DL or UL depending on the gNB’s scheduling decision and the transmission direction is indicated to the UE via a L1 control signaling. A slot consists of 14 symbols and 56 slot formats are defined to support various combinations of ‘D’, ‘U’, and ‘F’ symbols in a slot.
It can be seen that the dynamic TDD feature of NR supports more frequent and flexible switching between DL and UL direction in time than the LTE eIMTA.

%
%
%

For the CLI handling, two UE measurements were specified in Release 16 \cite{D5}: (1) SRS reference signal received power (SRS-RSRP) which can be utilized by gNB to estimate the interference from a UE’s uplink to another UE’s downlink; and (2) CLI received signal strength indicator (CLI-RSSI) which can be utilized by gNB to estimate the total amount of CLI experienced by a UE. 

\subsubsection{Duplex evolution in 5G-Advanced}
To further enhance the NR TDD operation in 5G-Advanced, 3GPP is performing a study on evolution of NR duplex operation in Release 18 \cite{D6}, which consists of two major subjects. The first subject is enhancement on the dynamic TDD to address, e.g., the gNB-to-gNB CLI, which may be caused due to either adjacent-channel CLI or co-channel CLI, or both, depending on the deployment scenario. The second subject is the feasibility of allowing the simultaneous existence of DL and UL, a.k.a., full duplex. 

\begin{figure*}[t]
\centering
	\includegraphics[width=0.9\textwidth]{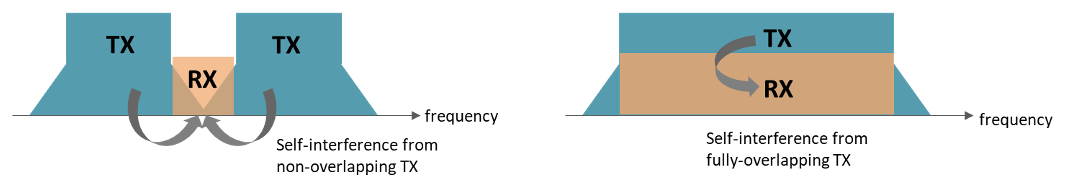}
	\caption{Self-interference from transmission to reception.}
	\label{fig:selfinterf}
\end{figure*}

Full duplex can create the CLI as observed with dynamic TDD as well as the self-interference from Tx to Rx. The significance of the self-interference depends on the frequency allocation for Tx and Rx as illustrated in Fig.~\ref{fig:selfinterf}, where the left case shows transmission and reception using the non-overlapping frequency while the right one shows transmission and reception using the overlapping frequency.
Considering the expected complexity for mitigating the self-interference from Tx to Rx and the CLI, the study assumes full duplex with non-overlapping allocation of DL and UL subbands at the gNB side while the UE performs the conventional TDD operation. Such approach could make the continuous UL transmission possible in the TDD system that is widely used in commercial NR deployments, which would be beneficial to improve coverage, latency, and capacity of the NR TDD system.

The non-ideal spectrum shaping and nonlinearity of power amplifier at Tx creates the self-interference and CLI to Rx even though DL and UL use different subbands. Separation of transmission and reception antennas possibly with a special structure for signal isolation between them and use of proper beamforming at Tx and Rx can achieve significant reduction of the self-interference. The remaining self-interference could be further suppressed by use of RF filters and digital signal processing at Rx. The Rx RF filter with sharp spectrum shape can also help reduce the CLI from adjacent carriers. On the other hand, the CLI caused by the transmission in the same carrier, especially DL transmission with high transmission power from a neighbour base station site to UL reception, would be much harder to mitigate. 
This property of CLI may make it easier to apply full duplex in the small cell deployment, where small cells (or small cell groups) are isolated so that the same carrier CLI can be easily avoided. 
For the wide-area macro cell deployment, proper coordination of transmission direction or UL-DL frequency separation between neighbouring base station sites would be necessary.

\subsection{Full Duplex Outside 3GPP}
Full duplex, also known as in-band full duplex (IBFD) where transmission and reception occur on the same frequency band, was first utilized as early as in 1940s in the context of full duplex radars \cite{D7}. Different from the FDD or TDD duplexing method when transmission and reception are sufficiently separated, full duplex imposes daunting challenges to transceiver implementations due to self-interference (i.e., a receiver is interfered by its own transmitter due to leakage) and potential CLI, as discussed earlier.

There are numerous comprehensive reviews on IBFD in the literature \cite{D8, D9, D10, D11, D12, D13}. Reference \cite{D8} focuses on full-duplex radars and IBFD wireless communications, presenting the corresponding opportunities and techniques for self-interference reduction. Reference \cite{D9} provides a comprehensive tutorial on IBFD from the perspective of physical and MAC layers. It summarizes the benefits, different topologies (i.e., bi-directional full duplex, full duplex relay, and full duplex cellular in a multiple access setting), challenges in self-interference cancellation, and IBFD for other purposes. Reference \cite{D10} compares more than 50 demonstrated IBFD communication systems with more than 80 different measurement scenarios, in terms of the corresponding isolation performance with respect to the center frequency, instantaneous bandwidth, and transmit power. A set of self-interference cancellation techniques are also summarized therein. Reference \cite{D11} presents a survey on IBFD in relaying scenarios, which covers the enabling technologies, key design issues, basics of IBFD, challenges and broader perspectives, and performance analysis. The tutorial in \cite{D12} concentrates on wireless communications, listing the potential benefits offered by full duplex techniques, surveying the critical issues related to full duplex transmissions from a physical-layer perspective relying on self-interference suppression/cancellation, while giving cognizance to the MAC-layer protocols, investigating the main hardware imperfections, and discussing the advantages, drawbacks, as well as design challenges of practical full duplex systems, and identifying their new directions and applications. Reference \cite{D13} contains a comprehensive survey on self-interference cancellation techniques in the antenna/propagation domain. It also discusses the opportunities and challenges of employing IBFD antennas in future wireless communication networks.

Besides radar and wireless communications particularly for 5G NR and 6G, full duplex also finds applications in many other areas, e.g.:

\begin{itemize}
    \item Cognitive radio networks, where full duplex operations enable simultaneous information sharing and sensing for improved spectrum sharing \cite{D14, D15};
    \item Physical layer secrecy, where full duplex helps activate more simultaneous transmissions or introduction of jamming noise signals, thus creating additional interference to eavesdroppers leading to the increased so-called secrecy capacity (i.e., the highest possible data rate at which a secret transmission can be reliably conveyed without being eavesdropped) \cite{D16};
    \item Wireless power transfer, where simultaneous wireless energy transfer (e.g., via DL) and wireless information transfer (e.g., via UL) can be supported, which is important for energy-constrained wireless communication systems (e.g., wireless sensor networks) \cite{D17}.
\end{itemize}

The challenges and potential solutions for IBFD were well studied in the literature \cite{D8, D9, D10, D11, D12, D13}. Effective and efficient self-interference and/or inter-node interference suppression/cancellation are critical and challenging for practical IBFD operations. There are many factors that need to be taken into account, e.g.:

\begin{itemize}
\item Time-varying channels; 
\item Single or multiple streams (i.e., MIMO); 
\item Potential presence of multiple transmitters and receivers (e.g., in cellular networks, where the issue is more pronounced in case of heterogenous networks); 
\item Hardware imperfections (e.g., phase noise, non-flat frequency response of circuits, power amplifier nonlinearity, transmit I/Q imbalance); 
\item Reasonable power consumption (especially for battery-powered devices) and complexity (e.g., to ensure small-size circuits). 
\end{itemize}

The enabling techniques for interference suppression/cancellation can be generally categorized into the following domains \cite{D10}:

\begin{itemize}
    \item Propagation domain, which may be comprised of passive, active and antenna interfaces;
    \item Analog domain, where the approaches may be in time-domain, frequency-domain, or digitally assisted;
    \item Digital domain, where various modelling may be assumed (linear/nonlinear, reference-based by utilizing an auxiliary receive channel, receiver beamforming, etc.). 
\end{itemize}

The study of the evolution of NR duplex operation in Release 18 in 3GPP marks an important step for practical IBFD operations. It is expected that the study would result in successful identification of the issues, benefits, and realistic solutions for IBFD, making it possible for commercial deployments in the future.

\section{ENERGY EFFICIENCY EVOLUTION TOWARDS GREEN NETWORKS}
\label{sec:green}

In this section, we focus on power consumption in wireless systems, which is one critical design aspect. Realizing green networks needs to consider both sides: UEs and base stations (network side). We first discuss UE side power consumption that may have direct impact on end-user experiences and thus has always been actively pursued and standardized in 3GPP. Then we consider the power consumption at the network side, which is crucial for network sustainability and vital for going forward.

\subsection{UE Side Power Consumption}
Power consumption for battery-powered end devices is an important performance metric. 3GPP standardization pays close attention to UE power consumption and has standardized various power saving techniques to improve UE power consumption, especially considering different UE types and operation conditions, even when there are conflicting goals.

5G NR is designed to support various UE types. Smart phone users are of course one primary target, which is often linked with the continued advancements on eMBB services \cite{G1}. These devices are generally highly capable with advanced processing power. MTC often aims to support low cost and low complexity devices, covering a wide range of applications such as metering, road security, and consumer electronic devices \cite{G2}. Different from smart phones, MTC devices generally require very long battery life, e.g., as long as 10 years \cite{G2}, since these devices may not be easily replaced due to the associated cost and operation environments. In order to more efficiently support services for connected industries, smart city innovations and wearables (e.g., smart watches), a new type of devices called RedCap devices \cite{G3}, with requirements lower than eMBB but higher than the traditional MTC, was recently standardized. 

Power savings for these devices are supported for different operation states, in particular, idle and connected states. An idle state UE is when the UE camps on a cell but without an established radio resource control (RRC) connection. Power saving techniques may include idle discontinuous reception (I-DRX) \cite{G2}, advanced paging techniques (e.g., group-based wake-up signal (WUS) for power savings \cite{G2}), relaxed measurements and reduced mobility requirements \cite{G3}. An RRC connected UE may benefit from connected state DRX (C-DRX), physcial downlink control channel (PDCCH) based WUS, and cross-slot scheduling (where a UE may perform the so-called “micro-sleep” if not scheduled \cite{G4}), restriction of low-rank MIMO operation when appropriate, reduced or skipped control channel monitoring, and fast activation and deactivation of a carrier or a configured scheduling. Monitoring PDCCH can be skipped or dynamically switched among a few configured search space set groups. A secondary cell may also be placed into a so-called dormancy mode as part of the activated state, where a UE is not required to monitor PDCCH for the secondary cell, but can be quickly indicated to transition from the dormancy mode to non-dormant mode and vice versa. A UE may also provide assistance information for power savings, such as a set of preferred power saving parameters including DRX configurations, a maximum aggregated bandwidth, and a number of carriers. For power savings, a UE can also have relaxed RRM, radio link monitoring, or beam failure detection requirements especially if its mobility level is low.

The goal for UE power savings may be in conflict with other design goals, e.g., capacity or coverage. As an example, an MTC user may need coverage enhancements as high as 20 dB \cite{G2}, which may require transmissions of the same information over an extended duration. Such repeated transmission due to the increased coverage need is not friendly to UE power consumption. Similarly, a UE with restricted MIMO operation is not capacity-friendly. 

More sustainable UE operations can benefit from energy harvesting,  especially for use cases such as health and fitness tracking, environment monitoring, and transportation \cite{G5, G6, G7, G8, G9, G10, G11, G12}. Energy harvesting may be particularly beneficial and associated with passive IoT devices such as RF identification (RFID) \cite{G10}. The mechanisms may be solar-based, wind-based, vibrational, electromagnetic, thermoelectric, or based on ambient RF energy \cite{G5, G12}. The challenges for RF energy harvesting may include overall RF power conversion efficiency, form factor, operational bandwidth, and compactness \cite{G7}. Reference \cite{G8} provides a summary of the RF energy harvesting in the past six years. 

\subsection{Network Side Power Consumption}
With the need to reduce the carbon dioxide (CO2) emissions and on the other hand to reduce the network operating cost, while facing continuous increase in data volume to be carried by the networks, improvements in the network energy efficiency is a must. Unlike the handset total energy consumption, the major part of the network energy consumption occurs during the lifetime of the network infrastructure instead of the actual manufacturing phase.  The development with 5G has already reduced especially the static power consumption compared to 4G or 3G, where networks had to be transmitting radio signals more frequently or even continuously even if there were no active data transmission taking place. With 5G the network side already took major step towards energy efficiency from continuous transmission with 3G or transmission every millisecond with 4G. This was also a requirement for the International Mobile Telecommunications-2020 (IMT-2020) technology \cite{G13}.

The key design choice with 5G was to move away from the common reference signals sent with such a frequency that one could not really drive down the power hungry RF parts even in case of lack of traffic. 3G networks were even worse with the common pilot channel which was sent continuously. The next steps in power efficiency on the network side are related to avoiding all transmissions when there is no actual user data to be transmitted, and using power efficient waveforms when having to transmit something. This together with other energy saving features allows 5G to be 10X more energy efficient than 4G in terms of energy consumed per traffic unit, but there still remains more that could be done as part of the 5G-Advanced. 

As part of the 3GPP 5G-Advanced studies \cite{G14}, the network energy consumption model is to be established, which then should enable vendor independent consideration of network power consumption impact when adding support for a new channel or signal to be transmitted or some new procedures between network elements. With the help of the energy consumption model, one is able to assess for example what would be the network side price to pay for a special signal to drive a low power wake-up radio at the UE side. This should be done to come up with sustainable designs for new features, which for the sake of saving on one side, would not cause major impacts on the energy efficiency in other dimensions.

The best energy saving is achieved when one can shut down an entire base station, especially first shutting down and replacing the earlier generations that consume lots of power regardless of the traffic level. Ideally, with only 5G remaining, one should optimize the number of frequencies to be active for the coverage and   traffic support needed at a given time in the area of interest, as well as the number of antenna ports active at the network side. Especially under low traffic load, the network could afford to use only a subset of antenna ports in the case of an antenna array, reducing the number of ports from, e.g., 64 to 16. AI/ML based control is expected to be the key enabler for this, allowing for dynamically coordinated energy saving operations across different network layers,  so as to minimize the network energy consumption in 5G \cite{G14}. As indicated in \cite{G14}, 70\% of operator energy consumption occurs at the base station site. Under  low traffic load, the possibilities are limited as the current solutions achieve most of the gains, and reducing the base station activity further would mostly make it even more  challenging to find and measure signals timely at the UE side, with diminishing gains as highlighted in \cite{G16}. 

The detailed solutions now part of the Release 18 scope for the network side aim to add operation without synchronization signal block (SSB), i.e., SSB-less, for secondary cell (SCell) operations for the inter-band CA case, as well as enhancing the cell discontinuous transmission (DTX)/DRX mechanism. For the spatial domain, adaptation of the number of spatial elements is to be enhanced. In the power domain, the power offset between physical downlink shared channel (PDSCH) and CSI-RS is to be adapted for lower energy consumption based on improved feedback to allow for more optimized PDSCH power setting at the gNB side. For the protocol side, conditional handover is to be enhanced to cover the case when either the source or the target cell is in the network energy saving mode.
 
Furthermore, there are various research directions on the improvements of  network energy consumption. One of the major factors worth to investigate is the impact of  network scheduling solutions. Besides the classical network performance, network power consumption is also impacted by the network scheduling approach, especially in 5G. In the case of low traffic load, ideally the scheduler would also send data only very sparsely, however  in 5G it is essential to pay attention to the resulting latency impact as well. The scheduling impact on network power consumption has been studied for example in \cite{G17} and in \cite{G18} for a given QoS level. Another dimension to consider is the use of heterogeneous networks with small and pico cells for improved energy efficiency to achieve the needed total capacity \cite{G19}. As small cells operate with lower power levels, they can offer capacity boost with good energy efficiency. In general, the use of self organizing networks (SONs) or lately AI/ML to obtain improvements in network energy efficiency has been discussed for example in \cite{G20}, and also in actual deployments, as there are many  network operation elements one can deploy without impact on the standards  \cite{G14}. It should be noted that when considering energy efficiency improvements, the energy cost for the enablers themselves such as AI/ML training and inference needs to be taken into account as well.

\section{ADDITIONAL 6G ASPECTS}
\label{sec:additional}

In the previous sections, we have discussed several 6G related aspects when introducing the key technology evolution directions in 5G-Advanced. The 6G discussions we have provided are by no means exhaustive. In this section, we point out some additional 6G aspects, including spectrum evolution, JCS for 6G, and provision of hyper-distributed, innovative 6G services.

\subsection{Spectrum Evolution}
To alleviate the scarcity of the spectrum as a precious resource, the concept of full spectrum is being propounded for 6G, which covers sub-7 GHz, centimetric range from 7-20 GHz, mmWave, THz, and optical frequency bands. 

The sub-7 GHz bands have been the dominant frequency spectrum in 4G and 5G, and will continue playing an important role in 6G, since they are naturally well suited to provide ubiquitous coverage and reliable wireless connection \cite{Spectrum1}. Nevertheless, the sub-7 GHz spectrum has been crowded and thus new spectrum is in need for future radio communications.

The pioneer spectrum blocks for 6G are expected to be in the centimetric range from 7-20 GHz for both higher data rates (as opposed to the sub-7 GHz band) and acceptable coverage, and the mmWave bands for peak data rates exceeding 100 Gbps. 
For example, the Federal Communications Commission (FCC) expressed the need to support 6G service in the mid-band spectrum such as the 7-16 GHz range and started to explore repurposing spectrum in the 12.7-13.25 GHz band for next-generation wireless technologies in the U.S.~\cite{FCC6GBand}.
Meanwhile, extensive research and trials have been carried out for the mmWave bands which have verified its great potential in cellular communications when combined with beamforming techniques. Thanks to its large bandwidths and small wavelengths (compared with lower frequency bands), mmWave is able to provide high temporal and spatial resolution, facilitating high-precision object localization and tracking. 

In addition, with the explosive growth of traffic globally, the THz spectrum is considered a promising new frequency range for 6G to provide even larger bandwidth and more abundant spectrum resources. In 6G wireless networks, typical application scenarios of THz communication may include indoor communication, hotspot downloading, wireless data center, fixed wireless access and wireless cellular fronthaul/backhaul, and security communication scenarios. However, based on the current state of technology, some critical challenges still need to be solved for THz communication to be applicable in 6G, including the corresponding superheterodyne transmission, modulators, channel models, channel estimation, beamforming and beam tracking, as well as signal generation and detection, antenna array manufacturing and efficiency, among others. 

Moreover, as a key driving force of the global Internet, the optical fiber network connects all continents to form a modern communication backbone network, providing high-speed data access for metropolises, cities, and towns. In some scenarios, it might be a promising option to extend optical fiber transmission directly to wireless interfaces to realize the connection and mobile access of the last mile. On the other hand, the visible light spectrum can be employed in LiDAR or other similar forms to sense the environment so as to assist radio communications, which is a popular use case of JCS.

\subsection{Joint Communication and Sensing}
\label{subsec:jcs}

JCS is a prospective field for 6G, which aims to realize the coordination of communication and sensing via software and hardware resource sharing or information sharing \cite{JCS1}. On one hand, in the beyond-5G era, the high frequency band used by wireless communication networks and the spectrum for radio sensing gradually approach or even overlap with each other. On the other hand, the communication system and the sensing system have similar features in terms of RF transceiver, channel characteristics, and signal processing, which hastens the research and development of JCS network architecture and related technologies. Moreover, the development of communication technologies, such as extreme large-scale arrays, large bandwidths, RISs, and AI, will further promote the growth of sensing technologies. 

In the narrow sense, sensing refers to target positioning (including ranging, velocity measurement, and angle measurement), imaging, detection, tracking, and recognition. In the broad sense, the sensing target may be services, networks, terminals, as well as attributes and states of environmental objects. JCS can be based upon the same set of equipment and/or the same spectrum, and is able to lower the equipment cost, size, and power consumption, and improve spectral efficiency, hardware resource utilization, and information processing efficiency. 

The development of JCS may face multi-perspective multi-level technical challenges, among which a few major challenges are listed below.
\begin{itemize}
\item \textbf{Fundamental theory for JCS}: Different from the classical Shannon information theory, sensing will give rise to new performance metrics and limits for the system, based on which new JCS information theory needs to be established to investigate the optimal performance limits and tradeoff of the two functionalities. 
\item \textbf{Signal processing for JCS}: This mainly embodies aspects such as joint waveform design, joint transmission beamforming, and joint signal reception. From the viewpoint of functionality priority, joint signal processing can be divided into sensing-centric design, communication-centric design, and joint weighted design.
\item \textbf{Protocol and system architecture design for JCS}: Communication and sensing may have different work mechanisms. Taking radar sensing as an example, radar is generally divided into pulse type and continuous-wave methods, while communication adopts TDD or FDD. Thus, new transmission protocol and system architecture are in need to realize the coordinated operation of the communication and sensing. 
\end{itemize}

The development of JCS is likely to undergo three stages, namely, coexistence, mutual assistance, and mutual benefit. For coexistence, the originally independent communication system and sensing system are integrated onto the same physical platform, where communication and sensing coexist as two service forms with the main focus on interference management, resource allocation, and spectrum sharing. In the mutual assistance stage, communication capability and sensing capability cooperate and aid each other based on shared information, so as to realize sensing-assisted communication or communication-assisted sensing, where the main focus lies in air interface design, waveform design, along with transmission and reception processing algorithms. Ultimately, in the most advanced stage, communication and sensing will achieve comprehensive and multi-level integration in spectrum resource, hardware equipment, waveform design, signal processing, protocol interface, networking collaboration, among other aspects. Besides the evolution of the existing techniques in the previous two stages, AI-enabled approaches and multi-cell coordinated sensing methods will be explored, in order to build the endogenous sensing ability of 6G.

\subsection{Hyper-distributed, Innovative Services}

Mission- and time-critical services will be essential to the further 
development of our societies and economy. Such services, like
connected healthcare, autonomous transportation, Industry 4.0, smart
grids, smart cities and homes, have strict QoS
requirements, which has led to new decentralized multi-tier
network-computing platforms where processing capabilities are brought
closer to the data sources and/or to the end consumers. This
multi-tier continuum, usually referred to as device-edge-cloud
continuum, thus describes integrated network-computing platforms that
consist of interconnected IoT/mobile devices, edge nodes, fog nodes,
and cloud back-end servers. In these platforms, computing resources
and services are seamlessly combined along the data path from the
IoT/mobile devices to the back-end clouds, passing through
intermediary edge and fog nodes.

Most of the aforementioned mission- and time-critical services
leverage ML for data analytics or decision making,
which makes it mandatory to support ML models, and the data pipelines to
feed them, in a sustainable manner. This concerns both the training of
such models, whose computing (and energy) resource requirements may be very 
high \cite{ML-energy},  as well as their execution for inference
purposes, especially when capability-constrained IoT/mobile devices
are used.

Additionally, it is worth remarking that among the most relevant services are media applications, which
represent over 80\% of the total Internet traffic. 
This suggests that human users' quality of experience should play a
central role in the design not only of media  applications, but also of 
the communication and computing networks supporting them.
It is thus of paramount importance to address jointly the definition of new media
and network solutions, in order to properly account for the users’
interaction and emotions, 
as well as the use of human digital twins to create an immersive experience.

In this context, 6G is expected to face some important challenges, as
set forth below.
\begin{itemize}
    
\item \textbf{Design and deployment of hyper-distributed services}: According to
the NFV concept, services are composed of several atomic virtual functions
chained together. 
This fact, combined with a multi-tier network and computing continuum and with the need to effectively support users' mobility, calls for novel strategies for service design and migration, where the  split into functions, their deployment, and their replacement upon users' movement can
be dynamically performed across 
the nodes in the continuum, so that the 
resources available therein are used at best and the service disruption time  is minimized.

\item \textbf{Intelligent services}: To make ML sustainable in spite of its
pervasiveness,  it is fundamental to enable collaborative ML model
building \cite{ADROIT} by leveraging, e.g., transfer learning techniques to reduce
computing and memory consumption. An ML model, locally trained in a
domain can be made available to other entities for further training and
update,  leading to continuous learning advancements, refinement and transfer across domains. 
Furthermore,  knowledge distillation or
pruning techniques can be smartly used to reduce the complexity of an
ML model without harming the quality of the decision-making process,
thus greatly reducing the amount of consumed resources.

\item \textbf{Semantic approaches}: To be able to transfer media traffic
enhanced with information related to the human behavior, emotions, and
personality without increasing the bandwidth demand exceedingly, it is
critical to develop 
synergic approaches that let services and networks closely
interact~\cite{CENTRIC}. More in general, being aware of the application and traffic
semantic can substantially help the network to identify, hence transfer and
process, only  those data that are essential and sufficient to creating a context and
providing a service that matches the user's preferences and needs, or the QoS required by an application. Initial solutions and future research directions in the area of semantic communications can be found in, e.g.,~\cite{infocom23-cp,semantic-petar,semantic-arxiv}.
\end{itemize}

\section{CONCLUSION} 
\label{sec:conclusion}

Over the past several years, we have witnessed a rapid deployment of commercial 5G networks that provide high-speed, low-latency connectivity for a wide range of use cases. New services with higher performance requirements will continue to emerge, calling for the continuous evolution of cellular technology. As this article has highlighted, 3GPP Release 18, the start of 5G-Advanced, includes a diverse set of evolutions that will significantly boost 5G performance and address a wide variety of new use cases. While we are just embarking on the 5G-Advanced journey, 6G research is already under way, and 6G standardization is expected to start within 3GPP around 2025. The innovative technology components investigated in 5G-Advanced evolution are essential precursors to key 6G building blocks, making 5G-Advanced a vital step in developing cellular technology towards 6G. 

6G will require even more data processing  at the edge of the network-computing
continuum for critical services, which can be achieved only with the tight cooperation of a
programmable network infrastructure supporting, primarily, end-to-end
network slicing across the multi-tier continuum with assured
QoS. Another aspect that raises unprecedented challenges to system
design is the presence, in many real-world vertical domains, of
interactive services based on distributed intelligence to support decision making. This requires that increasing amount of data, generated by massively deployed ubiquitous devices at the edge, is moved throughout the continuum to promptly build knowledge. On the other hand, a close inter-working between application and network transport layers offers the opportunity to develop  semantic approaches that enable the system to reconfigure according to the service to be supported, thus dramatically increasing efficiency, reducing energy consumption, and improving users' QoS.

It will be exciting to see how 5G-Advanced towards 6G will improve lives, foster industries, and transform society over the coming decade.


\section{Acknowledgment}
The authors thank Sai Sree Rayala (Virginia Tech) and Nima Mohammadi (Virginia Tech) for their help in typesetting the manuscript. 
This work was partially supported by NPRP-S 13th Cycle Grant No. NPRP13S-0205-200265 from the Qatar National Research Fund (a member of Qatar Foundation) and by U.S. National Science Foundation under grants CNS-2148212, ECCS-2128594, CNS-2003059, and CCF-1937487. The work of S. Sun is supported in part by the National Natural Science Foundation of China under Grant 62271310, and in part by the Fundamental Research Funds for the Central Universities of China. The findings herein reflect the work, and are solely the responsibility, of the authors.

\bibliography{IEEEabrv,reference}
\bibliographystyle{IEEEtran}
\end{document}